\documentclass[12pt]{article}
\usepackage{eurosym}
\usepackage[left=0.4in, right=0.4in, top=0.7in, bottom=0.7in]{geometry}
\usepackage{hyperref}
\usepackage{mathtools}
\usepackage{graphicx}
\usepackage{mathrsfs}
\usepackage{cite}

\newcommand{\ket}[1]{\bigl| #1 \bigr\rangle}
\newcommand{\subfigimg}[3][,]{%
	\setbox1=\hbox{\includegraphics[#1]{#3}}% Store image in box
	\leavevmode\rlap{\usebox1}% Print image
	\rlap{\hspace*{1.3cm}\raisebox{\dimexpr\ht1-1.1\baselineskip}{#2}}% Print label
	\phantom{\usebox1}% Insert appropriate spcing
}
\begin{document}
	
	\renewcommand{\baselinestretch}{1.3} \topmargin=-1.8cm \textheight=23 cm
	\textwidth=23cm
	
	\begin{center}

\textit{\Large  \bf{Detraction of decoherence that arises from acceleration process }\\ }

\bigskip	
		\textit{M. Y. Abd-Rabbou} $^{a}$ \textit{\footnote{E-mail:m.elmalky@azhar.edu.eg}},
		\textit{S. I. Ali}$^{a}$
	     \textit{\footnote{E-mail:salama5laser@azhar.edu.eg}},and  \textit{N. Metwally}$^{b,c}$
		\textit{\footnote{E-mail:nmetwally@gmail.com}}
		
		$^{a}${\footnotesize Mathematics Department, Faculty of Science, Al-Azhar
			University, Nasr City 11884, Cairo, Egypt.}
		
		$^{b}${\footnotesize Math. Dept., College of Science, University of Bahrain, Bahrain.}
		
		$^{c}${\footnotesize Department of Mathematics, Aswan University
			Aswan, Sahari 81528, Egypt.}
		
	\end{center}
	\begin{abstract}
	
The possibility of detracting the decoherence due to the acceleration process of the two-qutrit system is investigated, where we examined the behaviour of the relative entropy and the non-local information.  For this purpose,  the  accelerated  subsystems are  allowed to pass through local or global noisy channels.   It is shown that, the detraction potential depends on the type of the used noisy channel, local or global, and the initial settings of the  accelerated qutrit systems, whether it is  prepared in free or bound entangled intervals.  The improving rate that depicted for systems prepared in the free entangled intervals is much better than  those prepared in the bound entangled interval.  The maximum bounds of the non-local information in the presence of the amplitude damping channels are larger than those passes in the dephasing channel.

	\end{abstract}
\section{Introduction:}

Numerous studies have been dedicated to discussing the physical properties of quantum information processing between users.
 Different quantum systems are investigated to illustrate the influence of non-inertial frames on the degree of entanglement,  \cite{1,2}. The bipartite and tripartite entanglement of  the accelerated fermionic systems have been investigated, showing that as the acceleration increases, the entanglement decreases \cite{3}. The quantum correlations have received considerable attention from researchers in the presence of the Unruh effects, such as, the decoherence of a two-qubit system \cite{4},  the geometric measure of quantum discord for a two-qubit Hilbert space \cite{5}, the general behaviors of the quantum coherence for a single qubit \cite{6}, and  the dynamics of entropic uncertainty relation for two atoms \cite{7}. Another treatment that has been received much attention is the influence of noisy decoherence channels, which  may be considered as  the surrounding environments of the system. In general, it is realized that the quantum system suffers from decoherence because of the independent interaction between the system and these channels \cite{8}.  The influence of decoherence noise channel on the entangled states differs from that on both separable quantum systems and classical systems \cite{9}. The fragility of quantum entanglement via a simple two-qubit system has been investigated, which assumed dephasing noisy environments \cite{10,11}. The dynamical behaviours of quantum correlation for two-qutrit in the presence of global \cite{12}, local \cite{13}, multi-local noise, depolarizing noise \cite{14}, and amplitude damping \cite{15}  have been investigated.

It is well known that, the amount of  entanglement that may be generated between different quantum systems, represents the source of many quantum information tasks in the context of  quantum communication and computations \cite{16,17}. Experimentally,   entanglement  has been depicted for many quantum systems. For example, the entanglement  of two-atoms and four particles has been achieved using driving field \cite{18,19}. There are many measures of entanglement have been introduced as,  negativity, concurrence, tangled, and entanglement of formation \cite{20,21,22,17}.   On the other hand, The quantum coherence has different  mathematical structures to quantify the coherence quantum system \cite{23}.
 Among of these quantifiers is the relative entropy of coherence, where  it is  defined via von Neumann entropy \cite{24}. The relative coherence and the degree of entanglement are related to quantum superposition states, but this relation does not trust of the exact, where it is variant according to the dimension and surrounding environments \cite{25}. Indeed, to quantify the quantum coherence must have two essential elements: free states (i.e. separable states), and free operations \cite{23}. Besides the relative entropy of coherence, there are much coherence measures have been reconstructed, such as $ l_1 $ norm, trace distance measure \cite{26}, geometric measure \cite{27}, fidelity \cite{28}, and robustness of coherence \cite{29}.

Therefore we are motivated to consider  a system of two-qutirt, where it behaves as  a free entangled, bound entangled, or separable state depending on a weight parameter. It is assumed that both subsystems (qutrit) are uniformly accelerated. After tracing out the unobservable  degree of freedom of the accelerated system, the subsystem of the resulting state is compelled to pass through local or global noisy channels. The main task is investigating the behavior of different initial state settings free, bound, or separable state under the decoherence arises from the acceleration process and the noisy channels. We discussed the effect of these different noisy on the amount of quantum entanglement, and the decoherence.

 This paper is organized as: In Sec. (\ref{s1}), the suggested physical model and the acceleration process are introduced. The mathematical forms of the concurrence, relative entropy of quantum coherence, and non-local quantum information as a measure of the quantum correlations are reviewed in Sec. ( \ref{s2}).  It is assumed that, the accelerated subsystems interact independently with two different multi-local/global noisy channels. The effect of these noisy channels  on  some physical quantities is reported numerically in Sec.(\ref{s4}). Finally, Sec.(\ref{s5}) summarizes our results.

\section{The model:}\label{s1}
It is assumed that,  the  users Alice and Rob  share initially an entangled state of a two-qutrit system, which may be written as follows\cite{30},
\begin{align}\label{o}
	\hat{\rho}(\alpha)=\frac{2}{7}| \psi^+\rangle \langle \psi^+|+\frac{\alpha}{7} \sigma_+ +\frac{5-\alpha}{7} \sigma_-,
\end{align}
where $ \alpha$ is the setting state parameter. The seperable/entangled behavior of this system depends on the weight parameter $\alpha$, where $ \hat{\rho}(\alpha) $ is free entangled for $ \alpha \in (4,5]$, bound entangled for $ \alpha \in (3,4]$, and separable state for $ \alpha \in [2,3]$. The state vector  $ | \psi^+\rangle=\frac{1}{\sqrt{3}}(| 00\rangle+| 11\rangle+|22\rangle) $ represents the maximally entangled two-qutrit state. The operators   $ \sigma_+ $ and  $ \sigma_- $ are separable states, where $ \sigma_+ =\frac{1}{3}(| 01\rangle \langle 01|+| 12\rangle \langle 12|+| 20\rangle \langle 20|)$, $ \sigma_- =\frac{1}{3}(| 02\rangle \langle 02|+| 10\rangle \langle 10|+| 21\rangle \langle 21|)$.
It is  assumed that,  Alice's and Rob's qutrit are  accelerated simultaneously. Due to the acceleration process, the computational basis of the qutrit with k-mode $ \{\ket{0_k},\ket{1_k},\ket{2_k}\} $ in the Minkowski-Fock space are split up into two regions $I$  and $II$ in the Rindler-Fock space, which are given by \cite{31,32},
\begin{eqnarray}\label{2.5}
|0_k\rangle&=&\cos^2 r |0\rangle_{I} |0\rangle_{II}+e^{i \phi} \cos r \sin r \left(|1\rangle_{I} |2\rangle_{II} +|2\rangle_{I} |1\rangle_{II}\right)
\nonumber\\
&+& e^{2i \phi} \sin^2 r |\mathcal{P}\rangle_{I} |\mathcal{P}\rangle_{II},
\nonumber\\
|1_k\rangle&=& \cos r |1\rangle_{I} |0 \rangle_{II}+ e^{i \phi} \sin r |\mathcal{P}\rangle_{I} |1\rangle_{II},\nonumber\\
|2_k\rangle&=& \cos r |2\rangle_{I}|0\rangle_{II}-e^{i \phi} \sin r |\mathcal{P}\rangle_{I} |2\rangle_{II},
\end{eqnarray}
where $ |\mathcal{P}\rangle $ is the Minkowski of pair state, $r\in[0,\pi/4]$ is the acceleration parameter, and $\phi$ is an insignificant phase that can be omitted. By tracing out the degree of freedom on the second region $II$, one gets the  final density operator of the accelerated system on the first region $I$ as,
 \begin{align}\label{acc}
 	\hat{\rho}_{acc}(\alpha)&= e_{1,1}|00 \rangle \langle 00|+ e_{2,2}| 01\rangle \langle 01|+ e_{3,3}| 02\rangle \langle 02|+ e_{4,4}|10 \rangle \langle 10|+e_{5,5} | 11\rangle \langle 11|+ \nonumber\\ & e_{6,6}| 12\rangle \langle 12| + e_{7,7}|20 \rangle \langle 20|+e_{8,8} | 21\rangle \langle 21|+ e_{9,9}| 22\rangle \langle 22|+ e_{10,10} |0 \mathcal{P}\rangle \langle 0\mathcal{P}|+ \nonumber\\ &  e_{11,11} |1 \mathcal{P}\rangle \langle 1\mathcal{P}|+  e_{12,12}|2 \mathcal{P}\rangle \langle 2\mathcal{P}|+ e_{13,13} |\mathcal{P}0 \rangle \langle \mathcal{P}0|+e_{14,14} | \mathcal{P}1\rangle \langle \mathcal{P}1|+ e_{15,15}| \mathcal{P}2\rangle \langle \mathcal{P}2| \nonumber\\ &+e_{16,16} |\mathcal{P} \mathcal{P}\rangle \langle \mathcal{P}\mathcal{P}|+ e_{2,4}| 01\rangle \langle 10|+ e_{2,9}| 01\rangle \langle 22|+ e_{4,9}| 10\rangle \langle 22|+ e_{12,15}| 2 \mathcal{P} \rangle \langle \mathcal{P}2| \nonumber\\ & + e_{6,10}| 12 \rangle \langle 0 \mathcal{P}|+ e_{8,13}|21 \rangle \langle \mathcal{P} 0|+ e_{6,10}| 11 \rangle \langle  \mathcal{P}2|+ e_{8,13}|11 \rangle \langle 2\mathcal{P} |+h.c.
 \end{align}
where $ e_{i,j}~i,j=1...16 $ are a non-zero element for the accelerated system, with,
\begin{align}
	&e_{1,1}=\frac{\alpha}{21}\cos^8 r, \quad 	e_{2,2}=\frac{\cos^6 r}{21}\big(2+\alpha \sin^2 r\big)=e_{4,4}, \quad  e_{3,3}=\frac{\cos^6 r}{21}\big(5-\alpha \cos^2 r\big)=e_{7,7}, \nonumber\\ &   e_{10,10}=\frac{\cos^4 r \sin^2 r }{21}\big(7-\alpha \cos^2 r\big)=e_{13,13}, \  e_{6,6}=\frac{\cos^4 r }{21}\big(\alpha \cos^2 r+\sin^2 r (7+\alpha \sin^2 r)\big)=e_{8,8}, \nonumber\\ &  e_{11,11}=\frac{\cos^2 r \sin^2 r}{21}\big(\alpha \sin^4 r+(9-\alpha)\sin^2 r+5 \big)=e_{14,14},\	e_{2,4}=\frac{2\cos^6 r}{21}, \nonumber\\ &  e_{12,12}=\frac{\cos^2 r \sin^2 r}{21}\big(\alpha \sin^4 r+(12-2\alpha)\sin^2 r+\alpha+2 \big)=e_{15,15},\ e_{2,9}= \frac{2 \cos^5 r }{21}=e_{4,9},\nonumber\\ &
	e_{5,5}= \frac{\cos^4 r }{21}\big(\alpha \sin^4 r+4\sin^2 r-\alpha+5 \big), \ e_{9,9}= \frac{\cos^4 r }{21}\big(\alpha \sin^4 r+2(5-\alpha)\sin^2 r+2 \big),\nonumber\\ &  e_{16,16}=\frac{ \sin^4 r}{21}\big(\alpha \sin^4 r+(14-2\alpha) \sin^2 r+(7+\alpha) \big),\	\ e_{6,10}=\frac{2 \sin^2 r \cos^4 r}{21}=e_{8,13}, \nonumber\\ &  e_{12,15}=\frac{2 \sin^4 r \cos^2 r}{21},\  e_{8,13}=-\frac{2\cos^3 r \sin^2 r}{21}.
 \end{align}

\section{Information Measurements:}\label{s2}
In this section, we briefly familiarize the mathematical forms of the concurrence, the relative entropy of quantum coherence, and non-local information that encoded on the shared state between Alice and Rob.

\subsection{Concurrence:}
	Let us recall that the entanglement phenomenon via using the concurrence for $ m\otimes n $ mixed bipartite state (here $ m= n=3 $), which is defined by \cite{33},
\begin{equation}
	\mathcal{C}=Max\bigg\{0,\sqrt{\frac{2}{m(m-1)}}\big[\max\{||\rho^T||,||\rho^R||\}-1\big]\bigg\}
\end{equation}
here $ ||x|| $ is the trace norm of $ x $,  $ \rho_{ij,kl}^T=\rho_{kj,il}$, denotes a partial transposition of $ \rho $, and $ \rho_{ij,kl}^R=\rho_{ik,jl}$ is realigned matrix.
\subsection{Relative Entropy of Coherence:}
It is well known that, quantum coherence could be quantified by using relative von Neumann entropy \cite{23}.
It is   defined as the  minimum value of the  relative entropy between the set of all incoherent states $ I $ and the state $\rho$,
\begin{equation}
	\mathcal{R}_e=\min\limits_{\epsilon \in I}\mathcal{S}(\rho||\epsilon)=\mathcal{S}(\rho_{diag})-\mathcal{S}(\rho),
\end{equation}
where, $ \mathcal{S}(\bullet)=-Tr\{ \bullet \log_2 \bullet\} $ is the von Neumann entropy, $ \epsilon\in I $ with $ I $ is the set of all incoherent states, and  $ \rho_{diag}=\sum_{i} \rho_{ii}|i\rangle \langle i| $ is the diagonal matrix  of $ \rho $.

\subsection{Non-local Information.}
\quad The amount of quantum information that is  may be encoded  between  the subsystems is quantified via  the non-local quantum information\cite{34} $ \mathcal{I}_{\mathcal{N}}  $. In a mathematical form, it may be written as,
\begin{equation}\label{lo}
	\mathcal{I}_{non}=-Tr \rho \log_2 \rho =\sum_{i}\lambda_i^{ab} \log_2\lambda_i^{ab},
\end{equation}
where $ \lambda_i^{ab} $ are the real eigenvalues of the  Alice-Rob system.

\section{Numerical results}\label{s4}
In this section, we investigate the behavior of entanglement, the quantum coherence, and non-local through the concurrence $\mathcal{C}$, the relative entropy $\mathcal{R}_e$, and non-local von Neumann entropy. Different initial states settings of the two-qutrit system will be considered; free entangled, bound, and separable states. The effect of local/global noise channels on these quantifiers will be investigated numerically.

\subsection{Without noisy channels}
In this subsection, we investigate the  behavior of the amount of quantum correlation by means of the concurrence $\mathcal{C}$, where  we discuss the decoherence that arises from the acceleration process.
  \begin{figure}[h]
	\centering
	\includegraphics[width=0.35\linewidth, height=3.5cm]{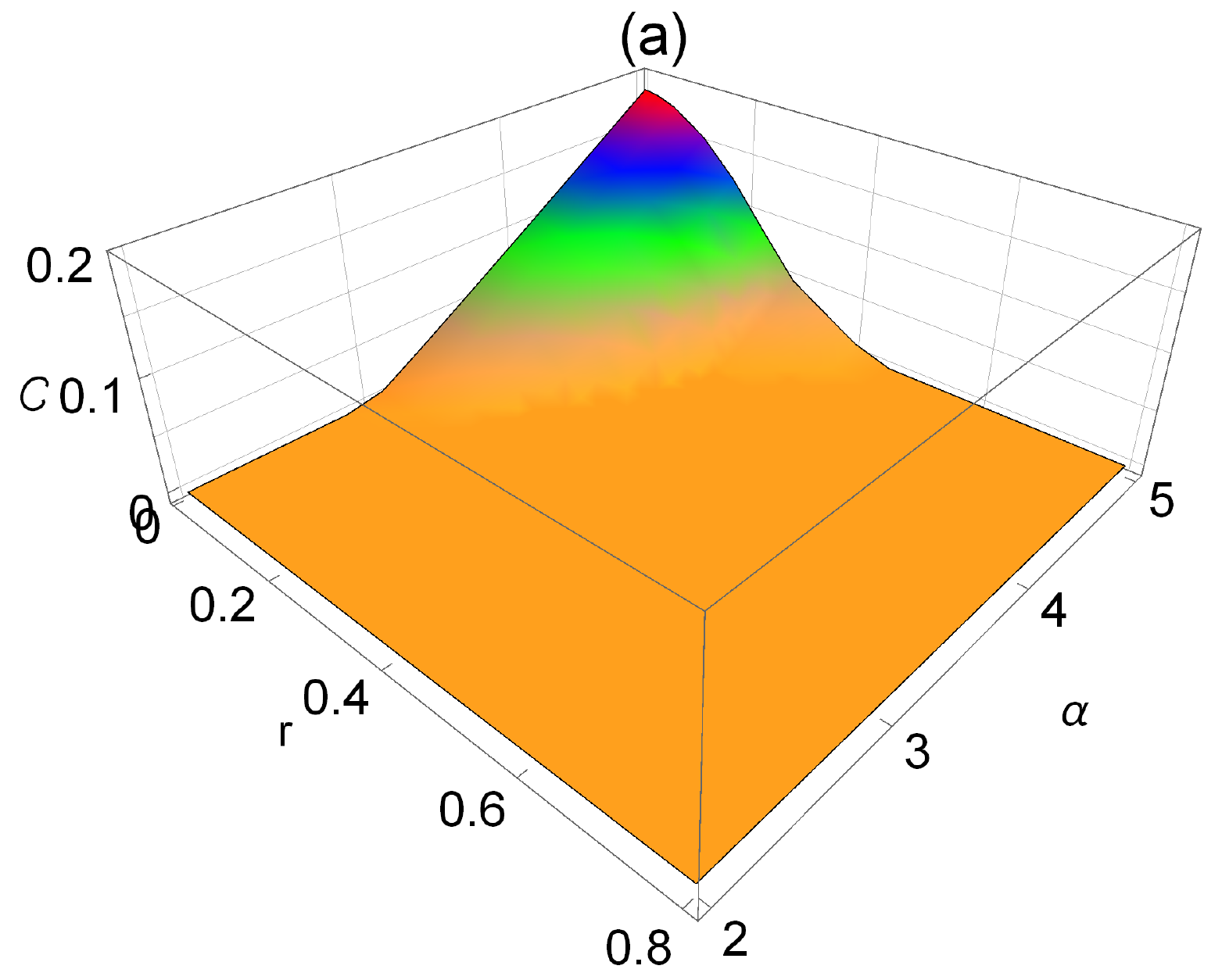} \hspace{0.2cm}
	\includegraphics[width=0.3\linewidth, height=3.5cm]{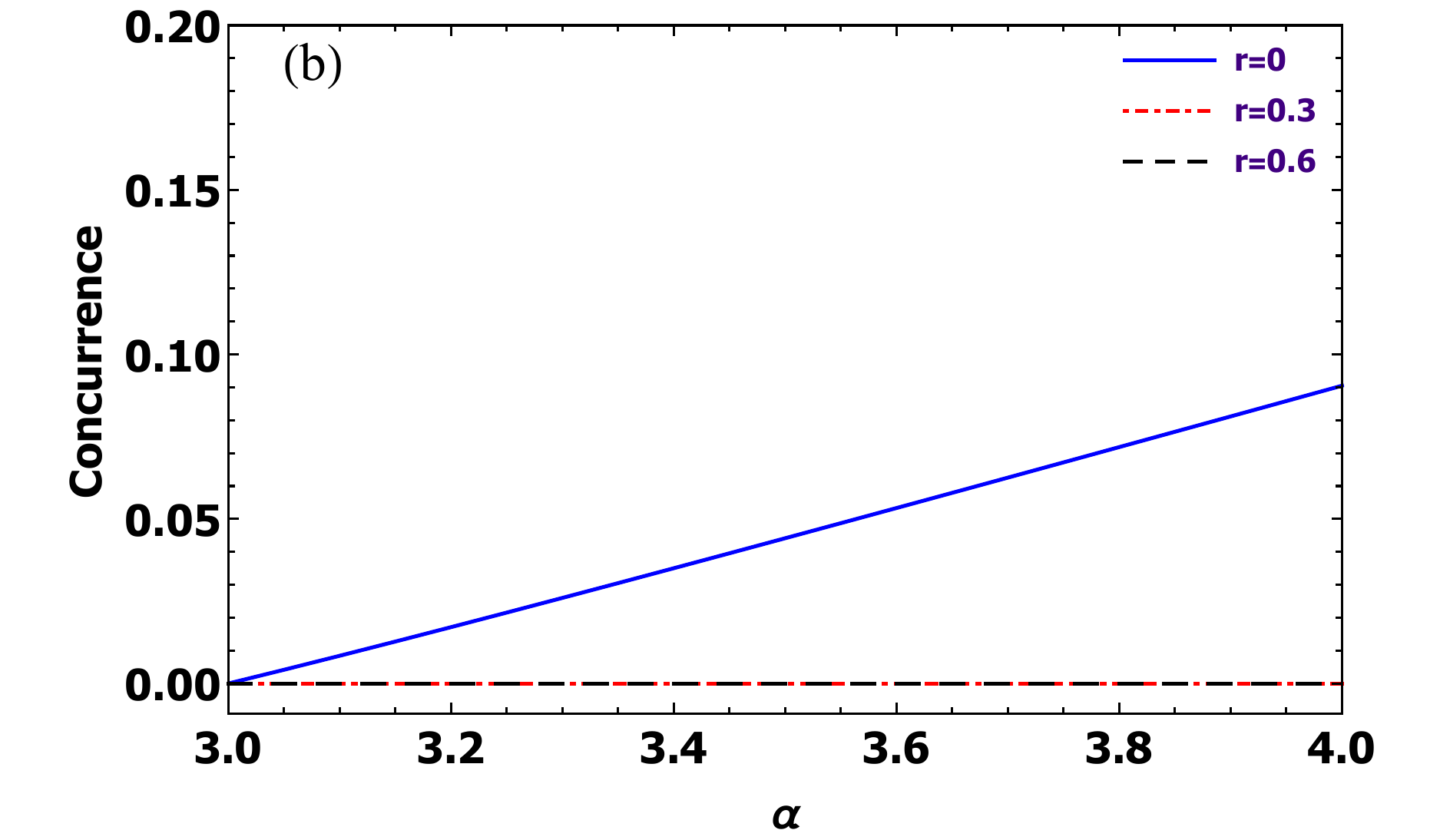}
	\includegraphics[width=0.3\linewidth, height=3.5cm]{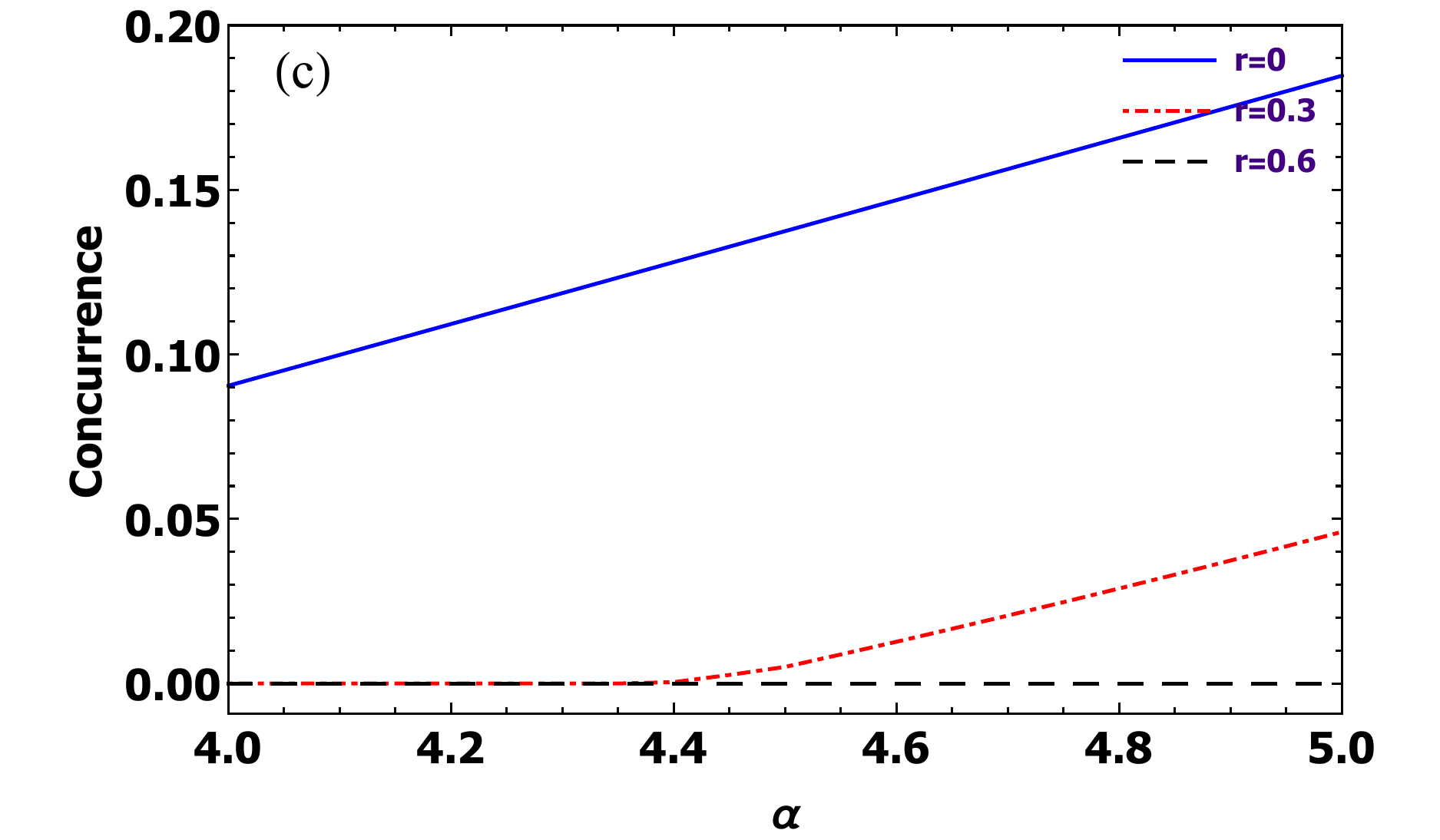}
	\caption{(a) The  concurrence $\mathcal{C}(r,\alpha)$ as a measure of entanglement, and (b),(c) represent $\mathcal{C}(\alpha)$  at different values of the acceleration, where the initial state  is prepared in  bound entangled and entangled intervals, respectively.}
	\label{fig1}
	\end{figure}
 In Fig.(\ref{fig1}), we investigate the behavior of  the concurrence, where all the possible settings of the initial state are considered, either separable, entangled or bound entangled. Fig.(\ref{fig1}a) shows the behavior of  concurrence $\mathcal{C}(r,\alpha)$, where there is no any quantum correlation when the qutrit system is initially  prepared in the separable interval, where $\alpha\in[2,~3]$, namely $\mathcal{C}(r,\alpha)=0$. However, the concurrence predicts the quantum correlation that could be generated between the subsystems, as the weight parameter $\alpha$ is chosen to be either on the entangled or bound entangled intervals.  Figs.(\ref{fig1}b) and (\ref{fig1}c) show the effect of the acceleration parameter on the behavior of the amount of the quantum correlation. It is clear that, the concurrence predicts the entanglement between the  two qutrits at  large acceleration, if the system is initially prepared  with $\alpha\in (4,5]$, namely on the free  entangled interval.
On the other hand, there is no quantum correlations are displayed   when the system is initially prepared  on the bound  entangled interval.

 \begin{figure}
	\centering
	
	\includegraphics[width=0.4\linewidth, height=4.5cm]{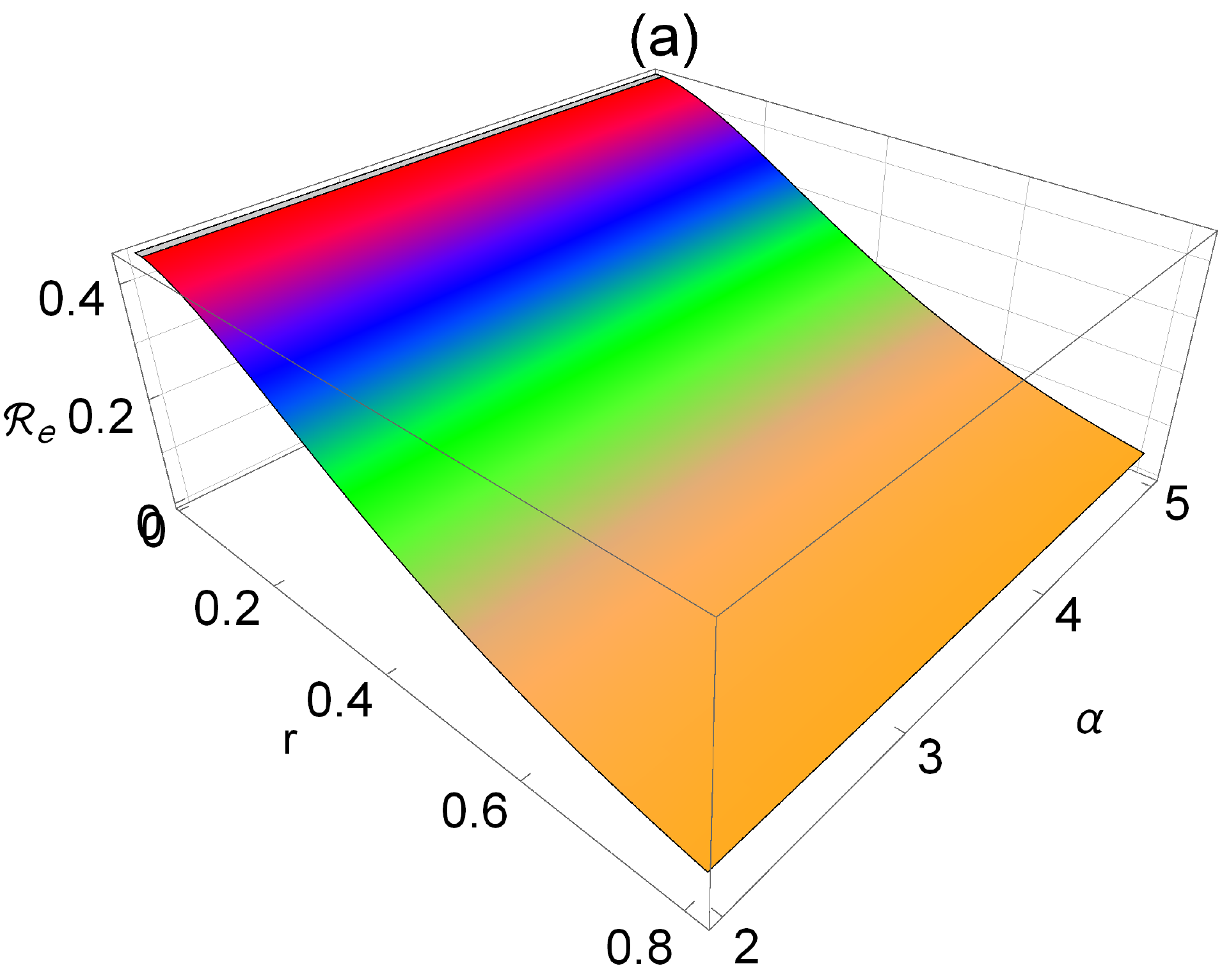}~~\quad
	\includegraphics[width=0.4\linewidth, height=4.5cm]{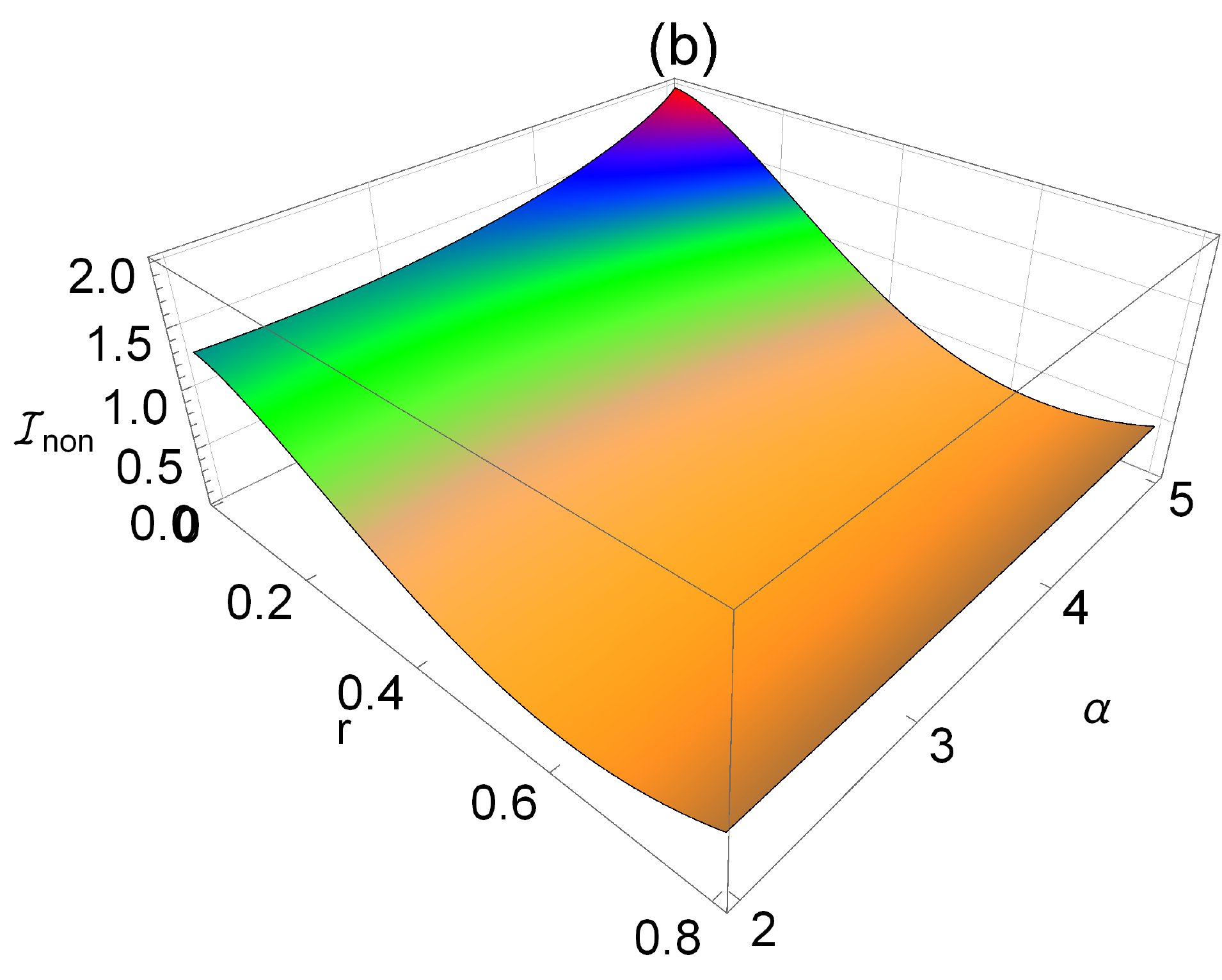}
	\caption{(a)The behavior of the relative entropy $\mathcal{R}_e$ as a measure of the  coherence, and  (b)the non local information, $\mathcal{I}_{non}$ of  the  accelerated system.}
\label{fig2}
\end{figure}

The effect of the acceleration on the coherence and the non-local information is displayed in Fig.(\ref{fig2}), where all the possible initial states settings of the parameter $\alpha$ are considered. It is clear that, both phenomena are decreased as the acceleration values increases. For the coherence, the decay depends only on the acceleration's parameter, while the non local information depends on the acceleration as well as on the initial state settings.

 \begin{figure}[h]
	\centering
	\includegraphics[width=0.42\linewidth, height=4cm]{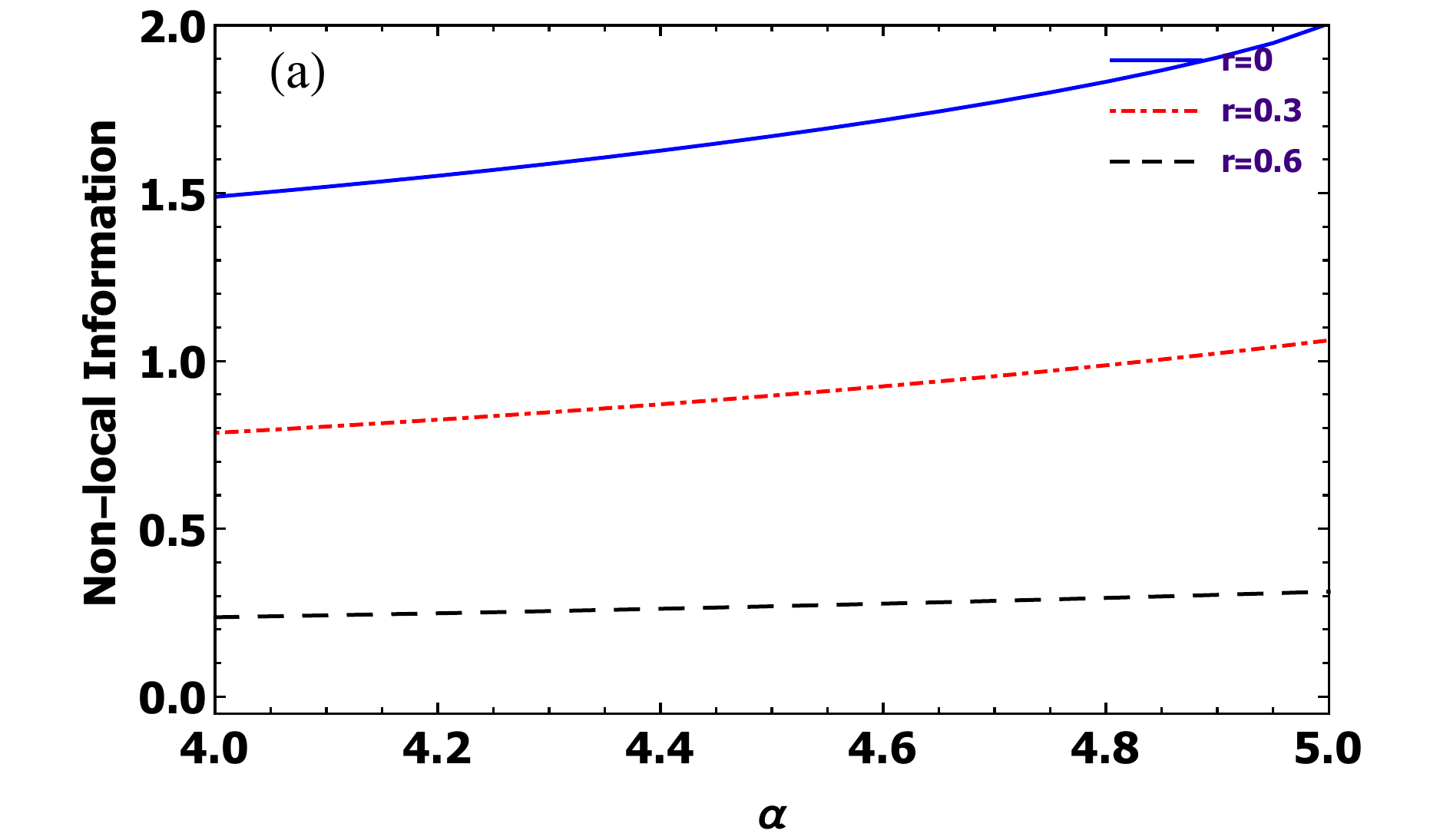}\hspace{0.2cm}
	\includegraphics[width=0.42\linewidth, height=4cm]{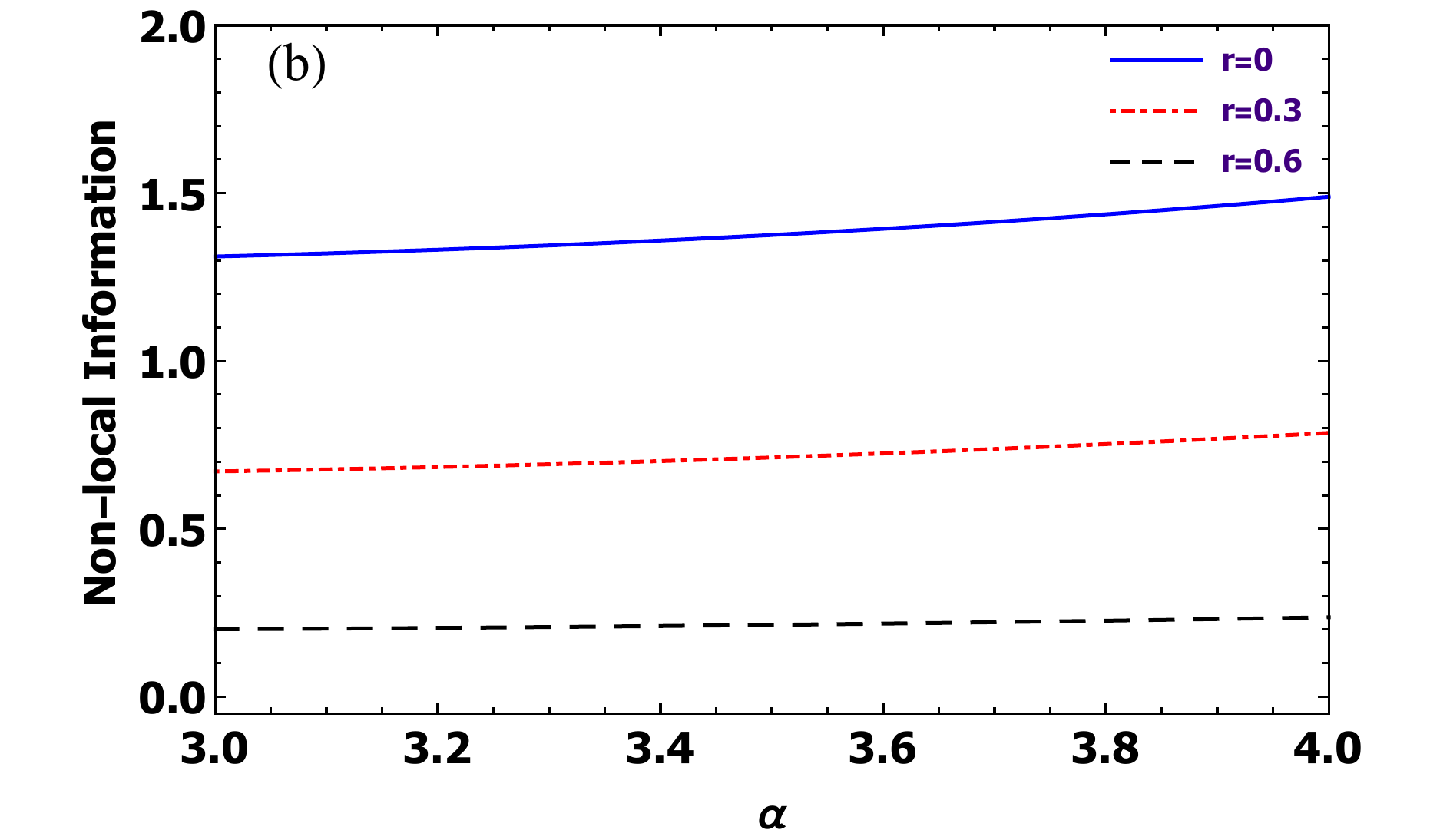}
	\caption{The behavior of the non local information, $\mathcal{I}_{non}$ where (a) the initial system is  prepared in the entangled interval, and (b) the system is initially prepared in the bounded entanglement interval.}
	\label{fig3}
	\end{figure}
As it is shown in Fig.(\ref{fig2}b) that the non-local information depends on the initial settings. Therefor it is important to investigate the behavior of $\mathcal{I}_{non}$ at some different values of the acceleration. Fig.(\ref{fig3}a) explores the behavior of $\mathcal{I}_{non}$, where  initial state is prepared such that $\alpha\in(4,5]$, namely on free entangled interval. In this case, the non local information increases gradually as $\alpha$ increases.
 Moreover, the acceleration has a passive effect, where the non-local information decreases as one increases the acceleration parameter $r$.  As it is displayed from Fig.{\ref{fig3}b), a similar behavior is displayed for states that prepared on the bound entangled  interval, i.e.  $\alpha\in (4,~5]$, but the increasing rate of the non-local information is smaller  than that displayed in Fig.(\ref{fig3}a).

From Figs.(\ref{fig2}) and (\ref{fig3}), one may conclude that, the decoherence that arises from the acceleration process destroys the amount of quantum correlation if the qutrit system is initially prepared in a bound entangled state. The decay phenomena  of the concurrence and the non-local information are displayed as the acceleration increases. However, the constancy behavior of relative entropy displays the coherence  depends slightly on the initial state settings, while it decreases as the acceleration increases.

\subsection{Within Noisy Channel:}\label{s3}
Now, we assume that both accelerated qutrit are interacted independently with their multi-local and global noisy decoherence channel.  In our treatment, we investigate the influence of the dephasing, and phase damping channels on the three  physical quantities  $\mathcal{C}$, $ \mathcal{R}_e $, and $\mathcal{I}_{non}$. If the final accelerated states (\ref{acc}) are compelled to pass through these noisy  channels, then the influence of multi-local decoherence operation is defined by\cite{35},
\begin{equation}\label{ch}
	\varrho_{m-ch}= \sum_{i,j}^{} \mathcal{E}_i^a \mathcal{E}_j^b \rho_{acc} (\mathcal{E}_i^a \mathcal{E}_j^b)^{\dagger},
\end{equation}
where $ \mathcal{E}_i $ are Kraus super-operator which are defined  below for each channel. Whereas, the final output state under the Kraus operator representation of global noisy decoherence channel is given by,
 \begin{equation}\label{chg}
 	\varrho_{g-ch}= \sum_{i,j,k}^{} \mathcal{E}_i^{ab} \mathcal{E}_j^{a} \mathcal{E}_k^b \rho_{acc} (\mathcal{E}_i^{ab} \mathcal{E}_j^a \mathcal{E}_k^b)^{\dagger},
 \end{equation}
with, $\mathcal{E}_j^{a}=\mathcal{E}_j \otimes I_3  $,  $\mathcal{E}_j^{b}= I_3 \otimes \mathcal{E}_j$, and $\mathcal{E}_j^{ab}=\mathcal{E}_j \otimes \mathcal{E}_j  $.

\subsubsection{Dephasing channel:}
It is assumed that, the accelerated system (\ref{acc}) is forced to path through the noisy channel, where the system interacted independently with multi-local dephasing decoherence channel and global noisy channel.  The Kraus super-operators of dephasing decoherence channel are defined by:
\begin{equation}\label{1}
	\begin{split}
		&\mathcal{E}_1=diag(1,\sqrt{1-\gamma},\sqrt{1-\gamma}) , \quad	\mathcal{E}_2= diag(0,\sqrt{\gamma},0) , \quad	\mathcal{E}_3= diag(0,0,\sqrt{\gamma}),
	\end{split}
\end{equation}
here $ \gamma $ is the quantum noise parameter.

\begin{figure}[h!]
	\centering
	\includegraphics[width=0.4\linewidth,height=4.5cm]{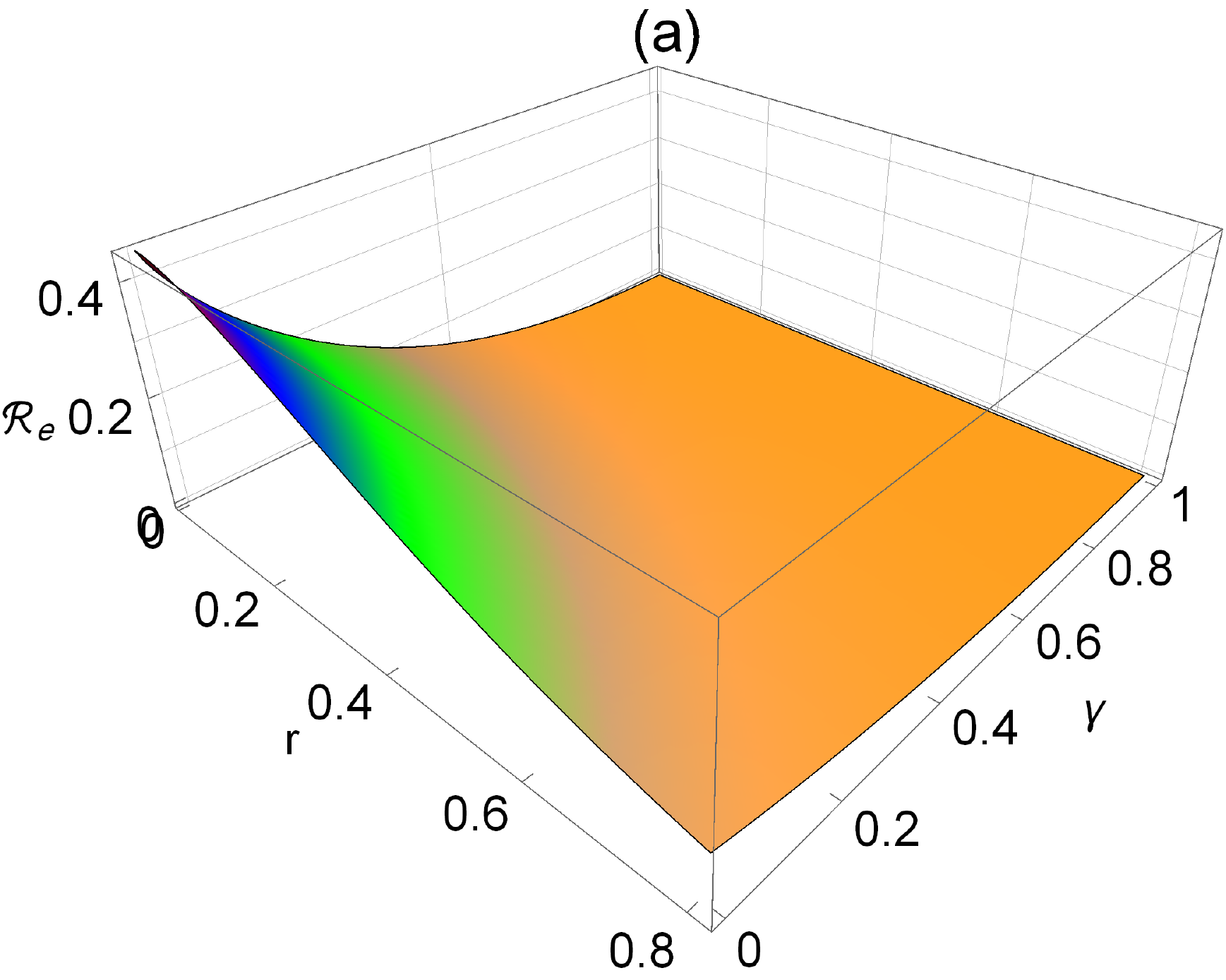}~\hspace{0.2cm}
	\includegraphics[width=0.4\linewidth, height=4.5cm]{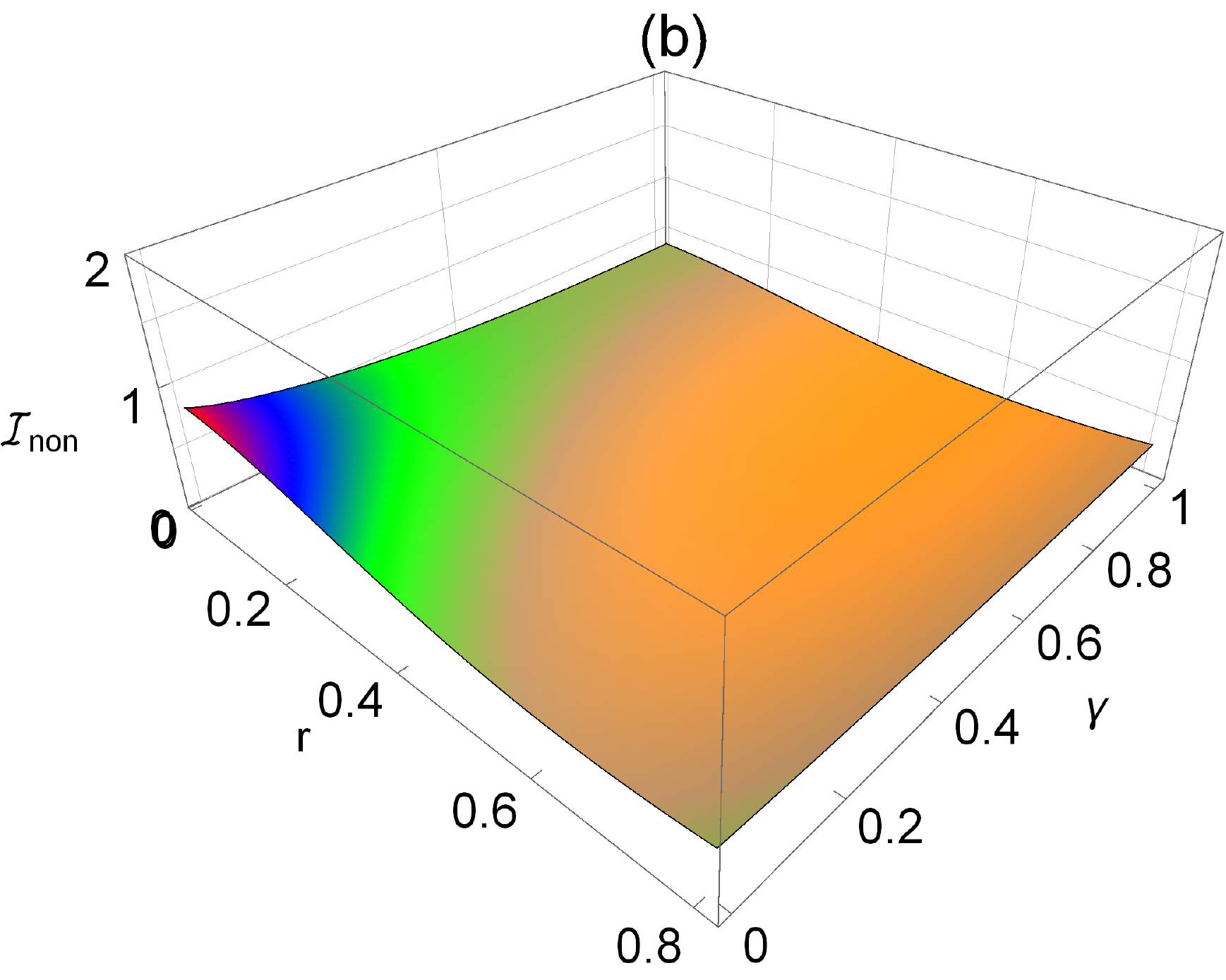}
	\includegraphics[width=0.4\linewidth, height=4.5cm]{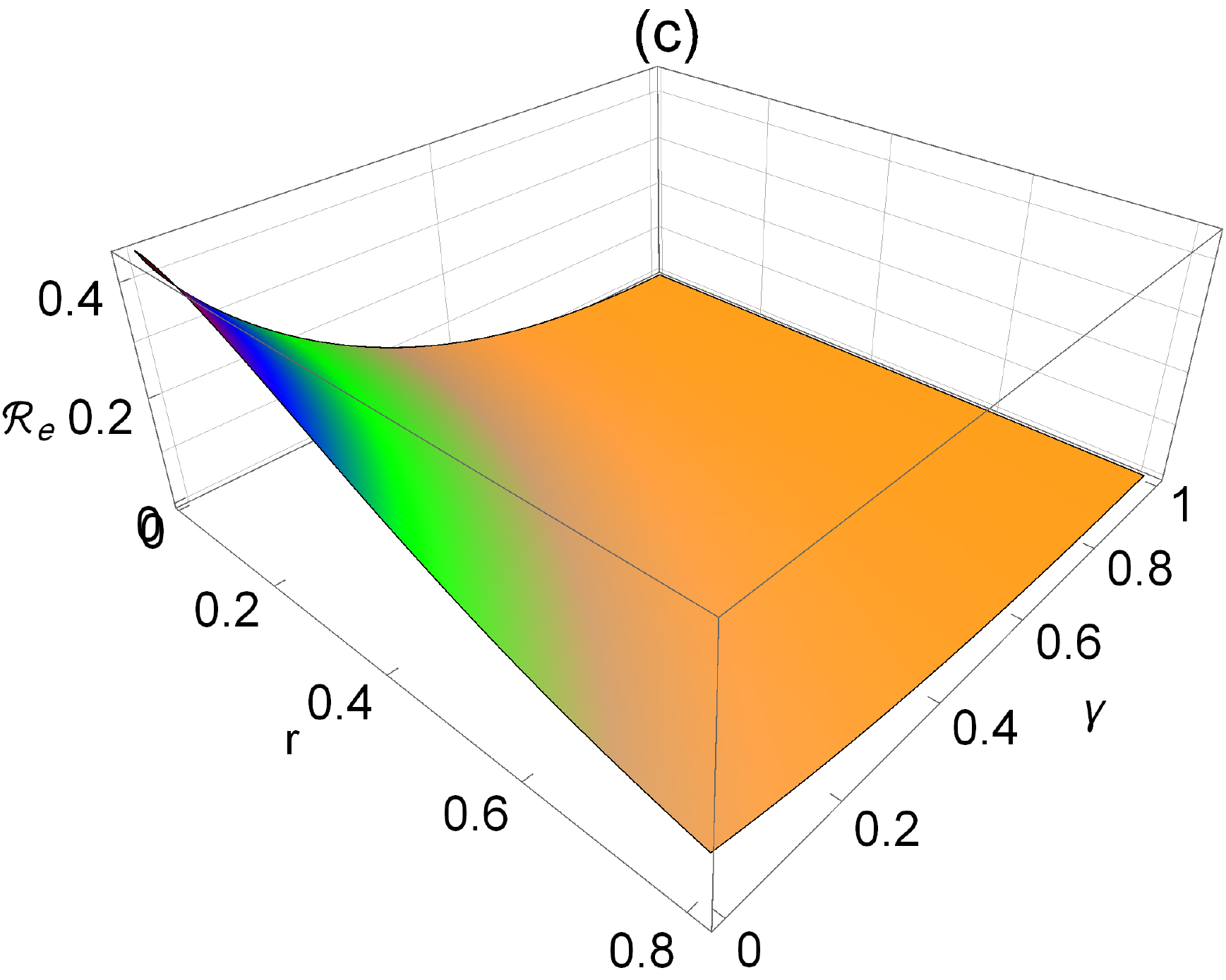}
\includegraphics[width=0.4\linewidth, height=4.5cm]{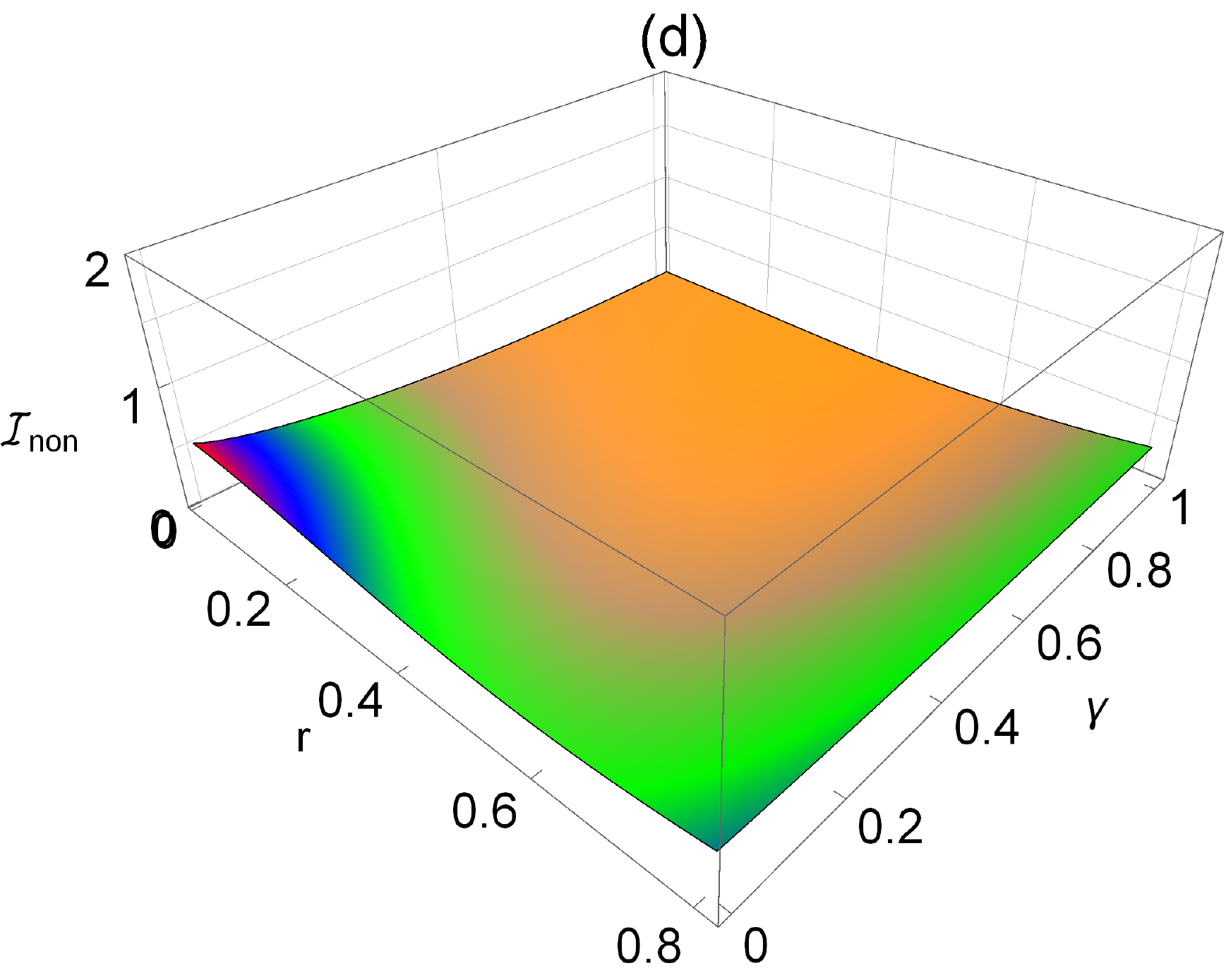}
	\caption{The influence of Multi-local dephasing channel on the accelerated system with $ \alpha=4.5 $ where, (a) the degree of  coherence, and (b) the non-local information. Figs.(c) and (d) are are the same as (a), (b), respectively but for a system is initially prepared in a bound entangled state with $\alpha=3.5$.}
	\label{fig4}
\end{figure}

In Fig.(\ref{fig4}), we discuss the effect of the multi-local dephasing channel on the behaviour of the degree of coherence and the amount of the non-local information, where  it is assumed that the system is initially prepared in the free entanglement interval, particularly we set  $\alpha=4.5$.
The depicted results show that the coherence $\mathcal{R}_e$ and the non-local information $\mathcal{I}_{non}$ decreases gradually as the acceleration parameter $r$ and the channel strength $\gamma$ increase.  As it is shown from Fig.(\ref{fig4}b), the free entanglement state can robust the decoherence due to the channel strength $\gamma$, where as $r\to \infty$, the non-local information $\mathcal{I}_{non}$ doesn't vanish at any value of $\gamma\in[0,~1]$. On the other hand, the coherence almost vanishes either at the maximum values of the channel strength or at maximum values of both  decoherence parameters $r$ and $\gamma$. Moreover, the decoherence due to the acceleration process is larger than that arises from the  channel strength. In Figs.(\ref{fig4}c} and (\ref{fig4}d), we investigated the behavior of $\mathcal{R}_e$ and $\mathcal{I}_{non}$ for an initial qutrit system prepared in a bound entangled interval, where we set $\alpha=3.5$.  The behavior of both quantifiers is similar to that displayed for free entangled system (Figs. (\ref{fig4}a) and (\ref{fig4}b)), but the upper bounds are smaller. Moreover, the maximum/minimum values of the non-local information have appeared at different  intervals of the entangled and bound entangled qutrit states. This observation is clear by comparing Fig.(\ref{fig4}b) and (\ref{fig4}d), where for the free entangled qutrit system, the maximum values of $\mathcal{I}_{non}$ are displayed at $r\simeq\in[0.~~0.8]\bigcup[0.6,~0.8], \gamma\in[0,~0.2]\bigcup [0,~~1]$. While for the bound entangled qutrit system, the maximum values of $\mathcal{I}_{non}$ are depicted on $r\simeq\in[0.~0.2]\bigcup[0.0,~0.8], \gamma\in[0,~1]\bigcup [0.8,~1]$.

\begin{figure}[h!]
	\centering
	\begin{tabular}{@{}p{0.43\linewidth}@{\quad}p{0.43\linewidth}@{}}
		\subfigimg[width=\linewidth, height=3.7cm]{\textbf{(a)}}{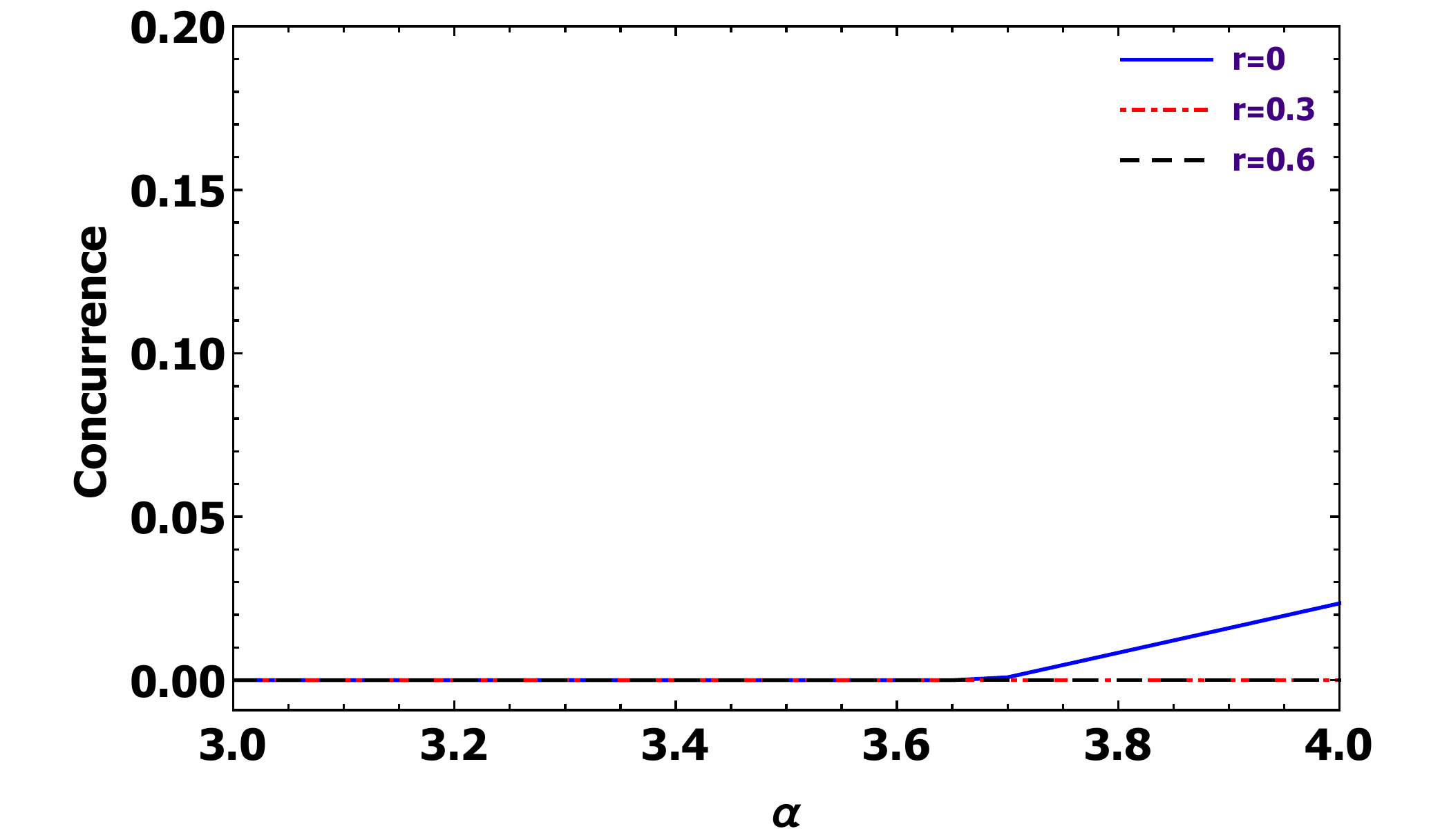} &
		\subfigimg[width=\linewidth, height=3.7cm]{\textbf{(b)}}{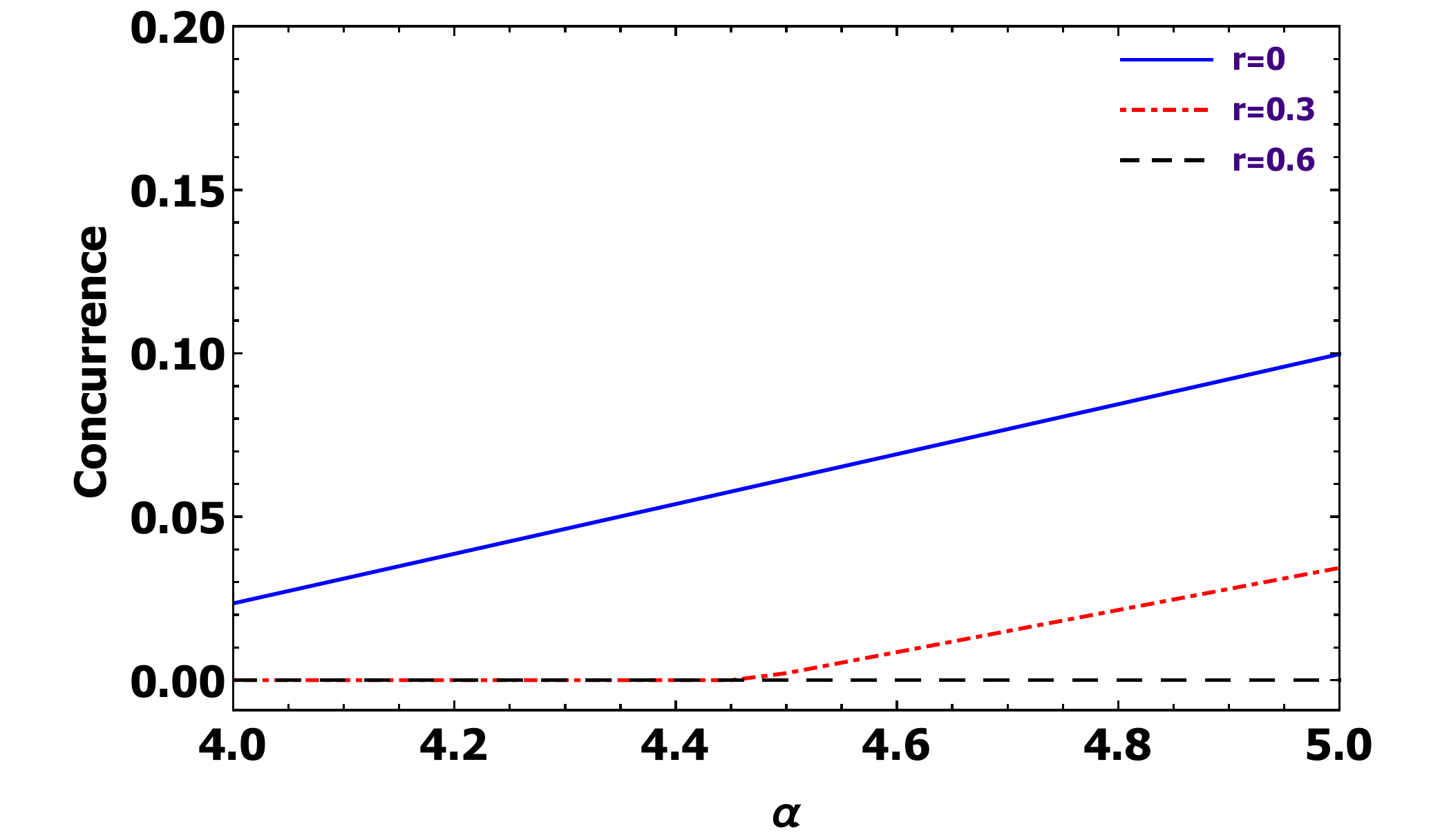} \\
		\subfigimg[width=\linewidth, height=3.7cm]{\textbf{(c)}}{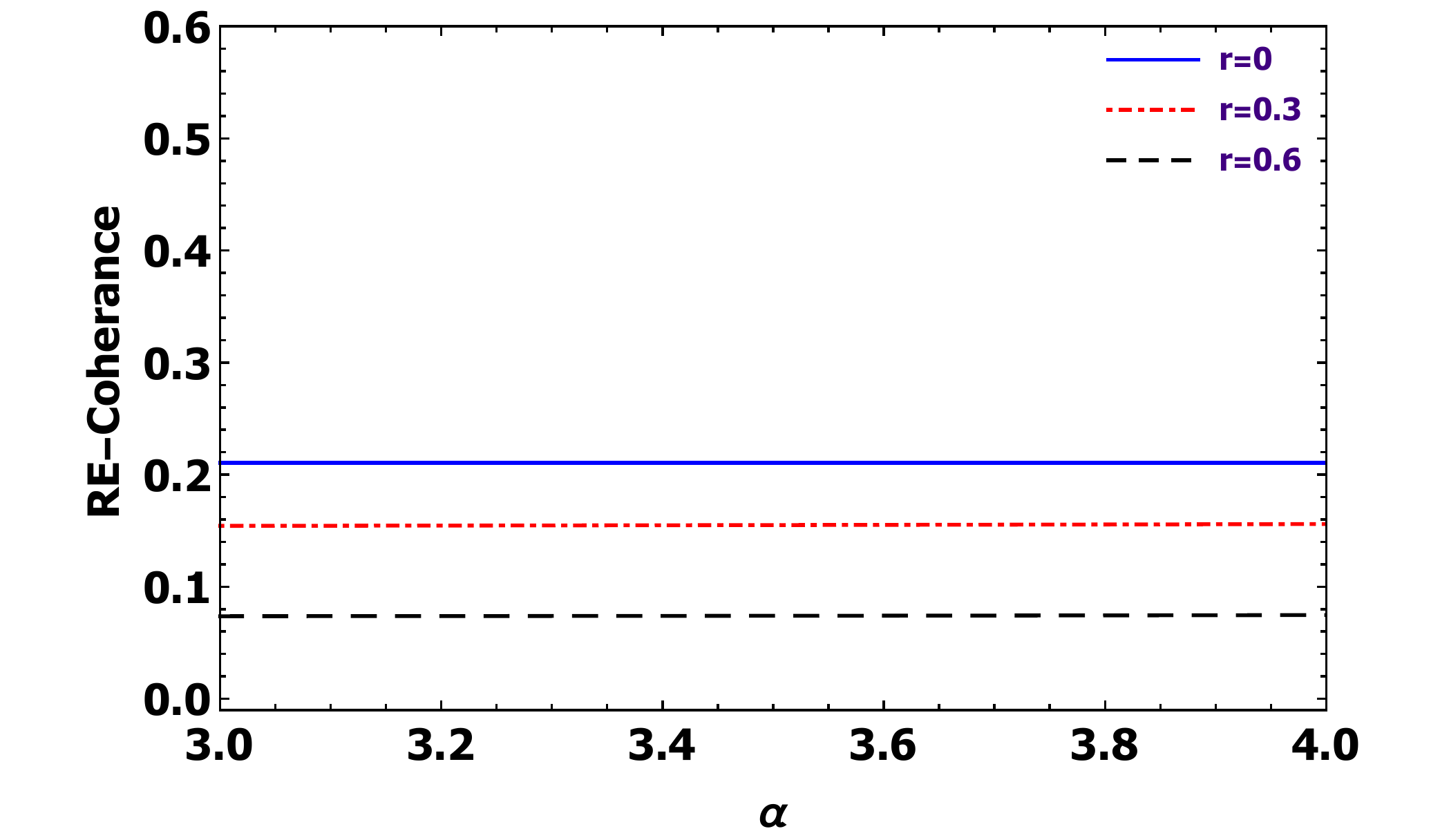} &
		\subfigimg[width=\linewidth, height=3.7cm]{\textbf{(d)}}{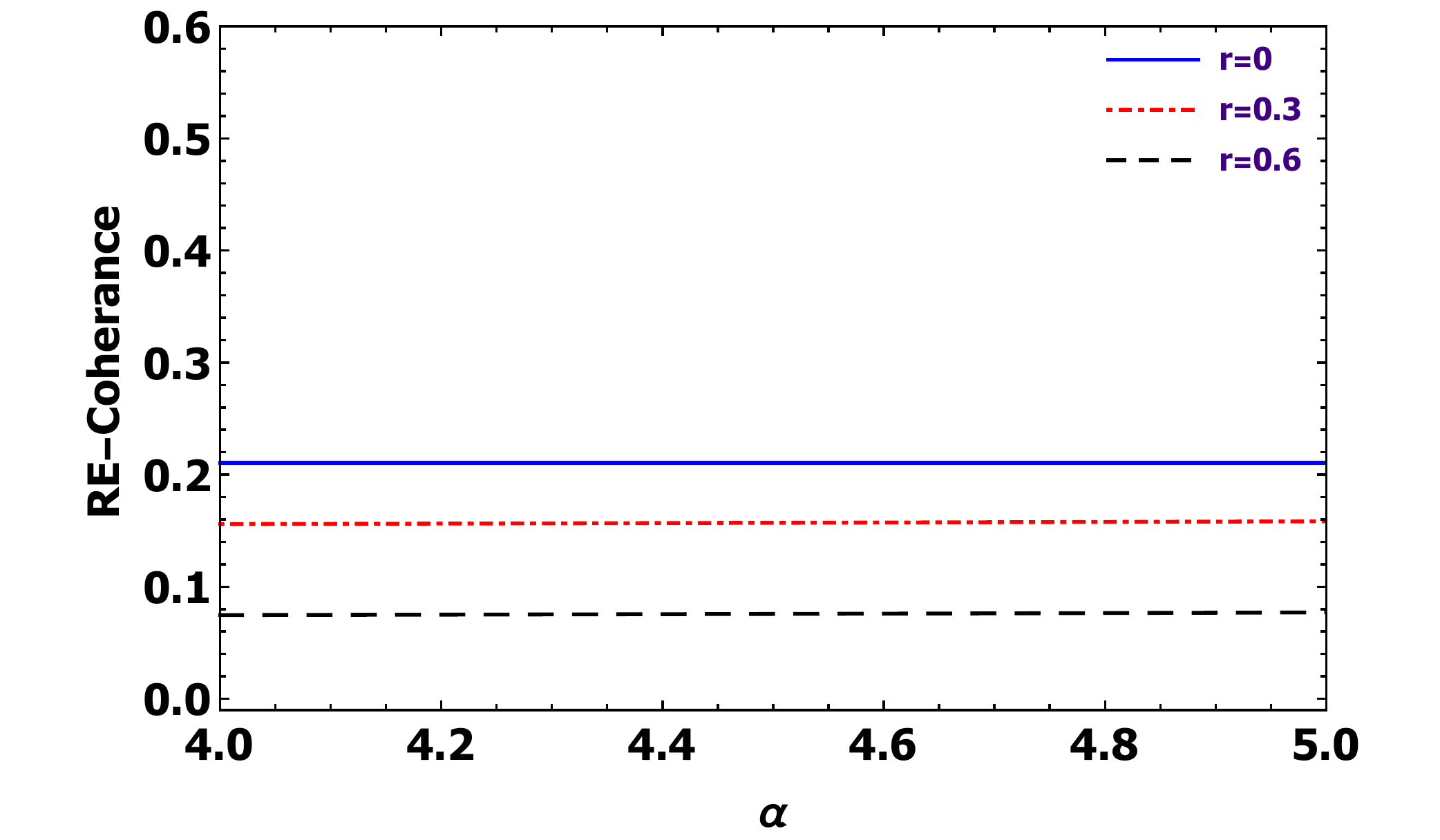} \\
		\subfigimg[width=\linewidth, height=3.7cm]{\textbf{(e)}}{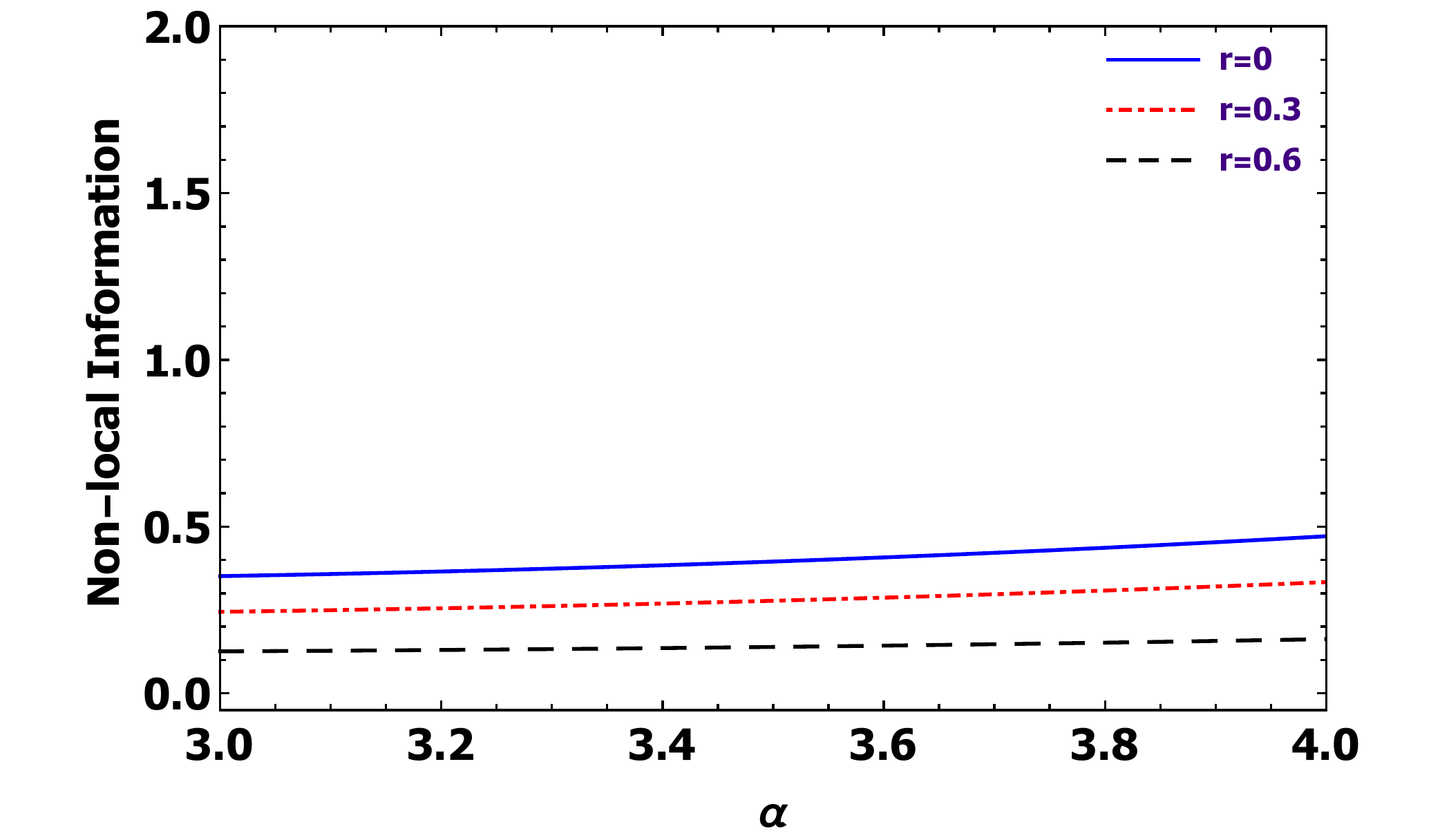} &
		\subfigimg[width=\linewidth, height=3.7cm]{\textbf{(f)}}{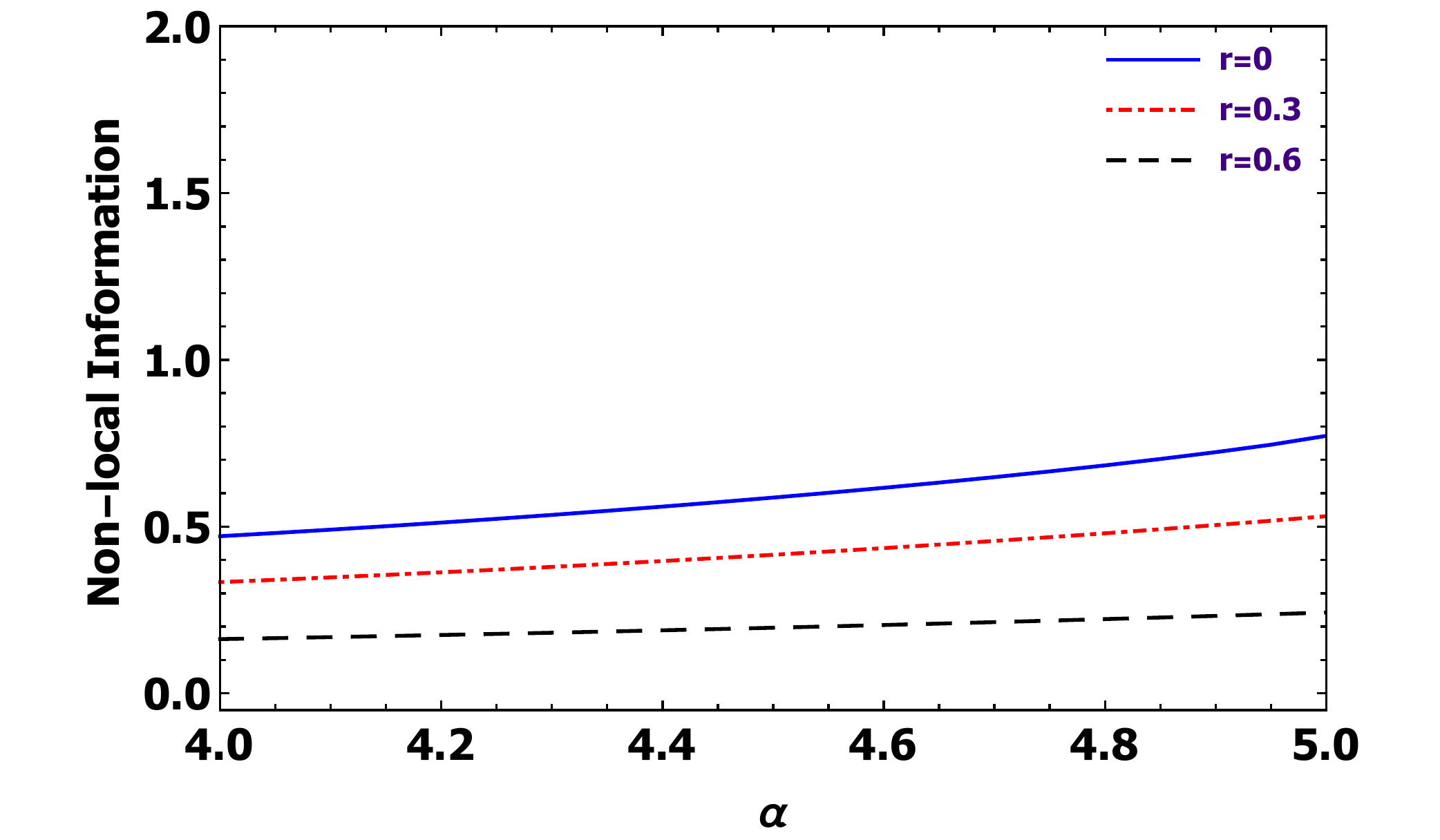}
	\end{tabular}
	\caption{Influence of multi-local amplitude damping channel with $ \gamma=0.1 $ (a,b) Concurrence, (c,d) Coherence, and (e,f)Non-local information.}
	\label{fig5}
\end{figure}

The behavior of the three quantities, $\mathcal{C}$, $\mathcal{R}_e$ and $\mathcal{I}_{non}$ at some different values of the acceleration and $\gamma=0.1$ is shown in Fig.(\ref{fig5}) for free and bound entangled initial states. It is clear that, the upper bounds of all the three quantities are smaller than that displayed in Figs.(\ref{fig1}) and (\ref{fig3}). At large acceleration ($r=0.3)$, the quantum correlation and the non-local information are displayed at larger values of the weight parameter $\alpha$.  The decreasing rate that predicted for free entangled is smaller than that displayed for bound entangled states. The maximum bounds of the coherence  are  smaller than that displayed in the absence of the local noisy channel.

\begin{figure}[h!]
	\centering
	\includegraphics[width=0.4\linewidth, height=3.7cm]{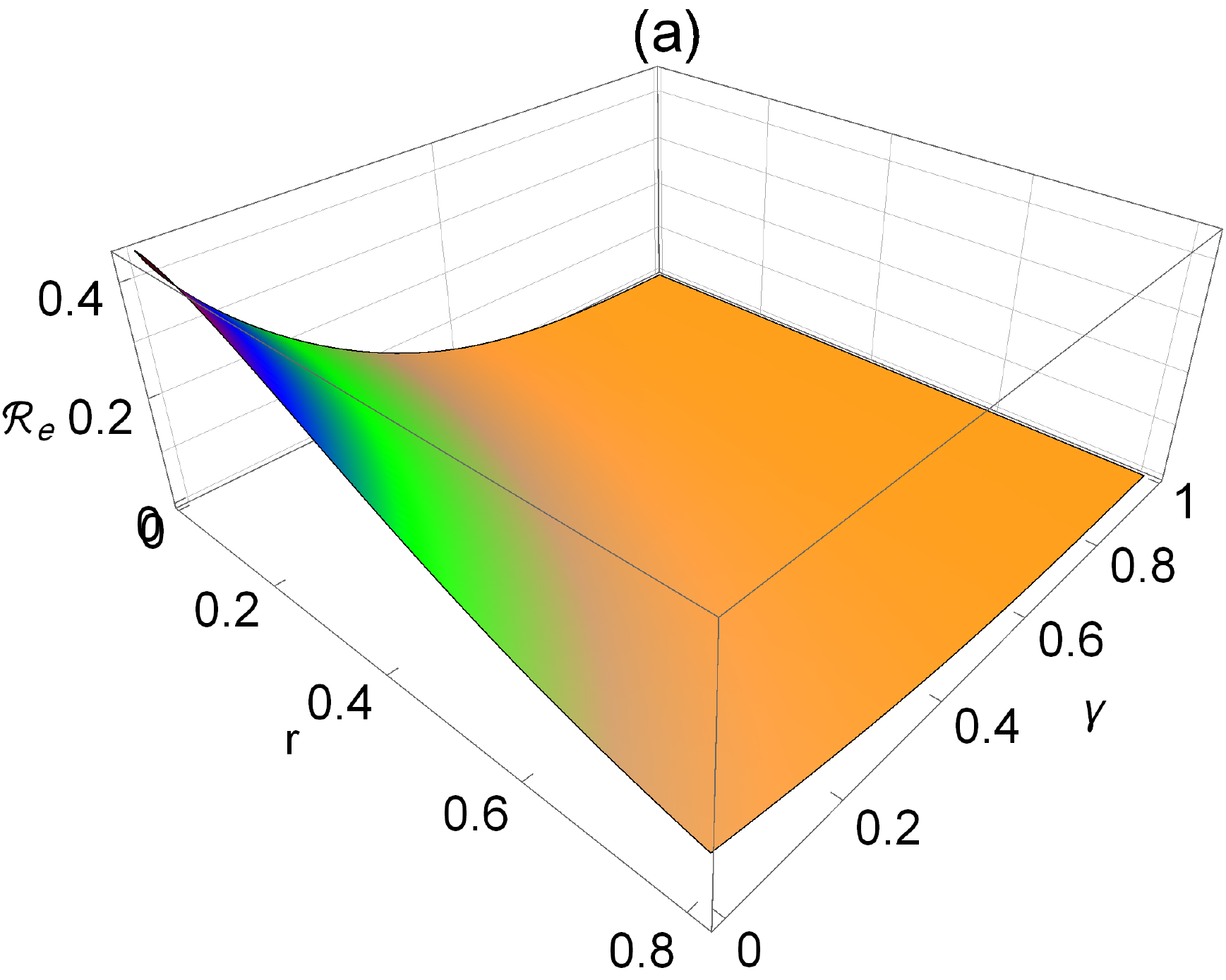} \hspace{0.2cm}
	\includegraphics[width=0.4\linewidth, height=3.7cm]{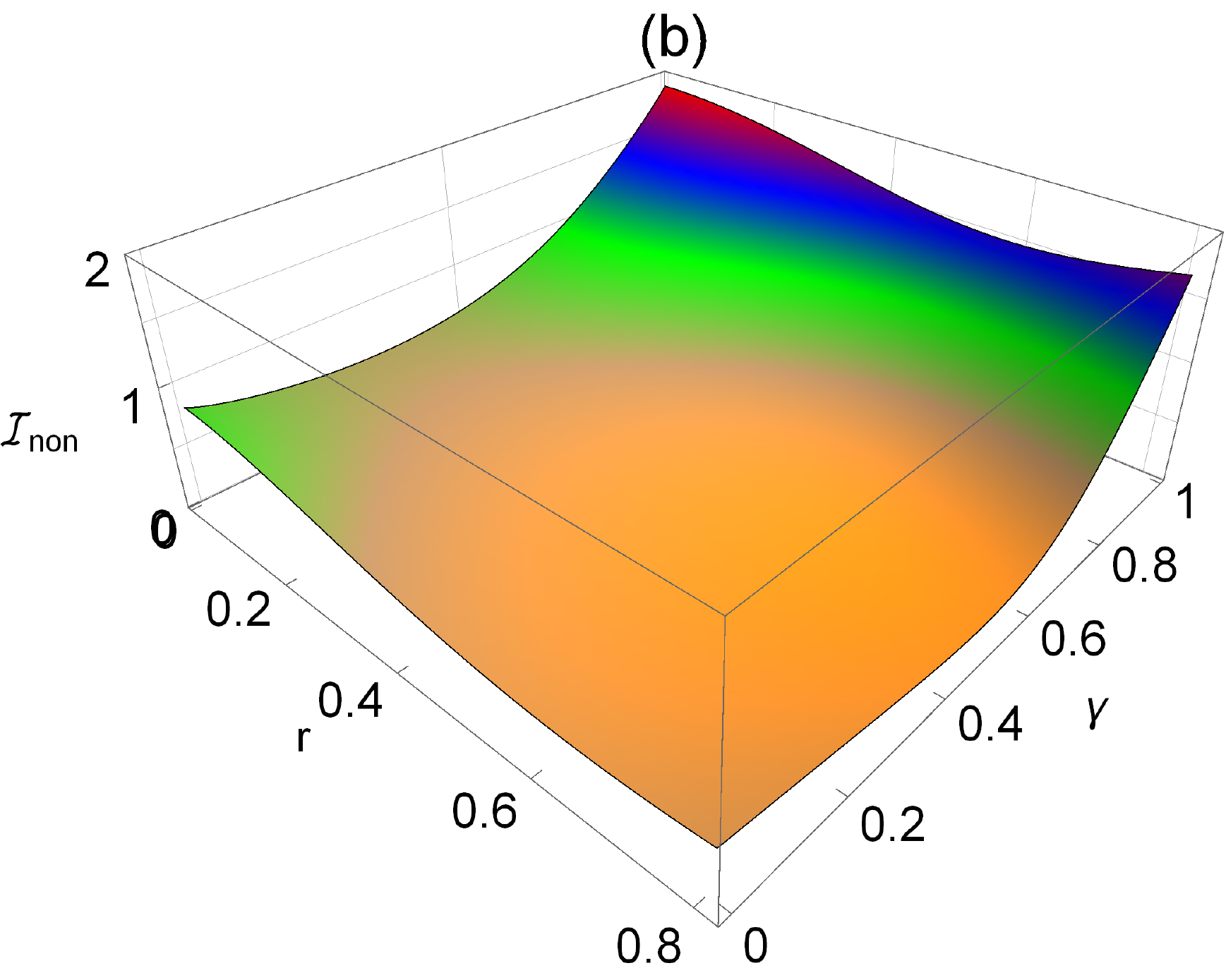}\\
	\includegraphics[width=0.4\linewidth, height=3.7cm]{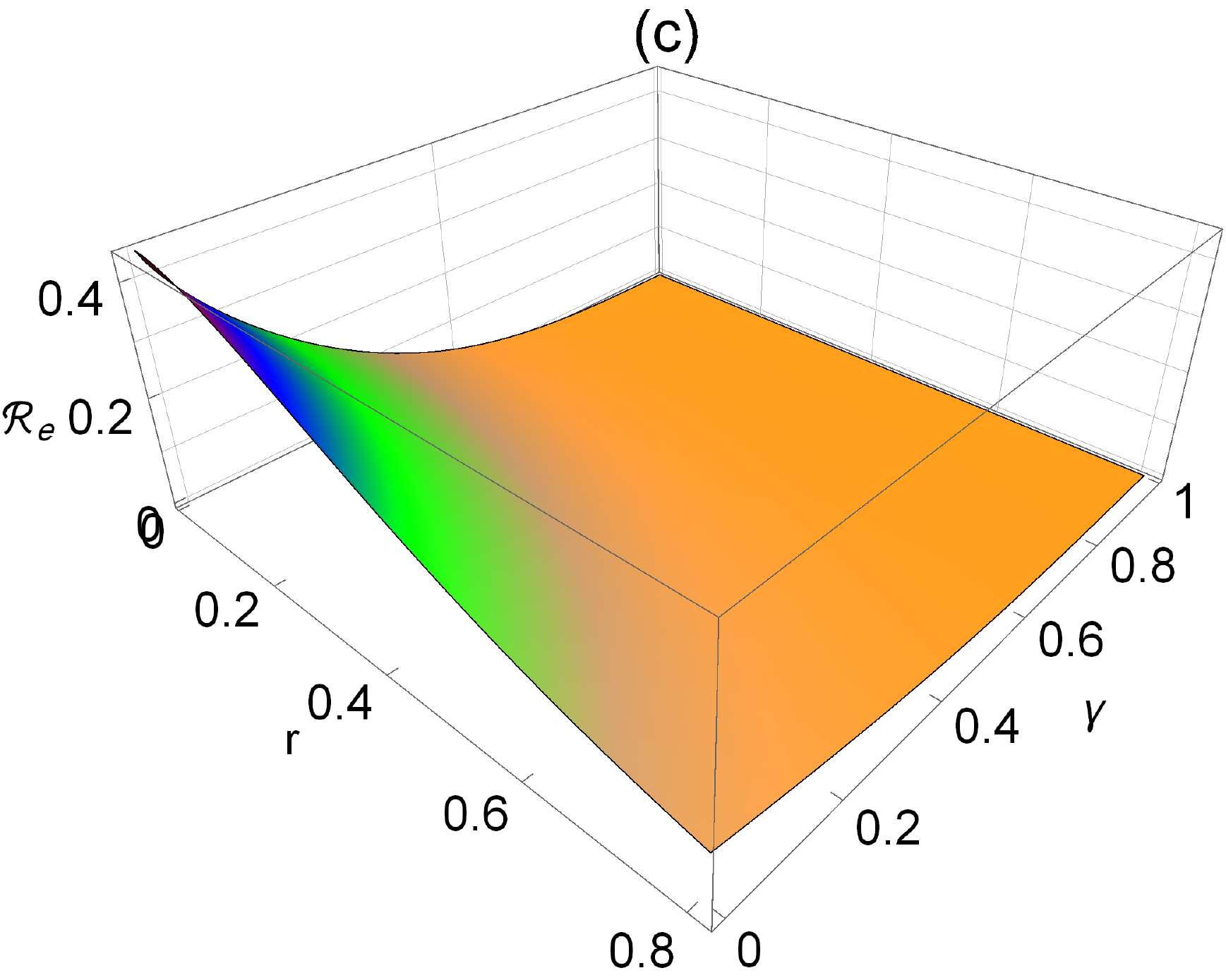} \hspace{0.2cm}
	\includegraphics[width=0.4\linewidth, height=3.7cm]{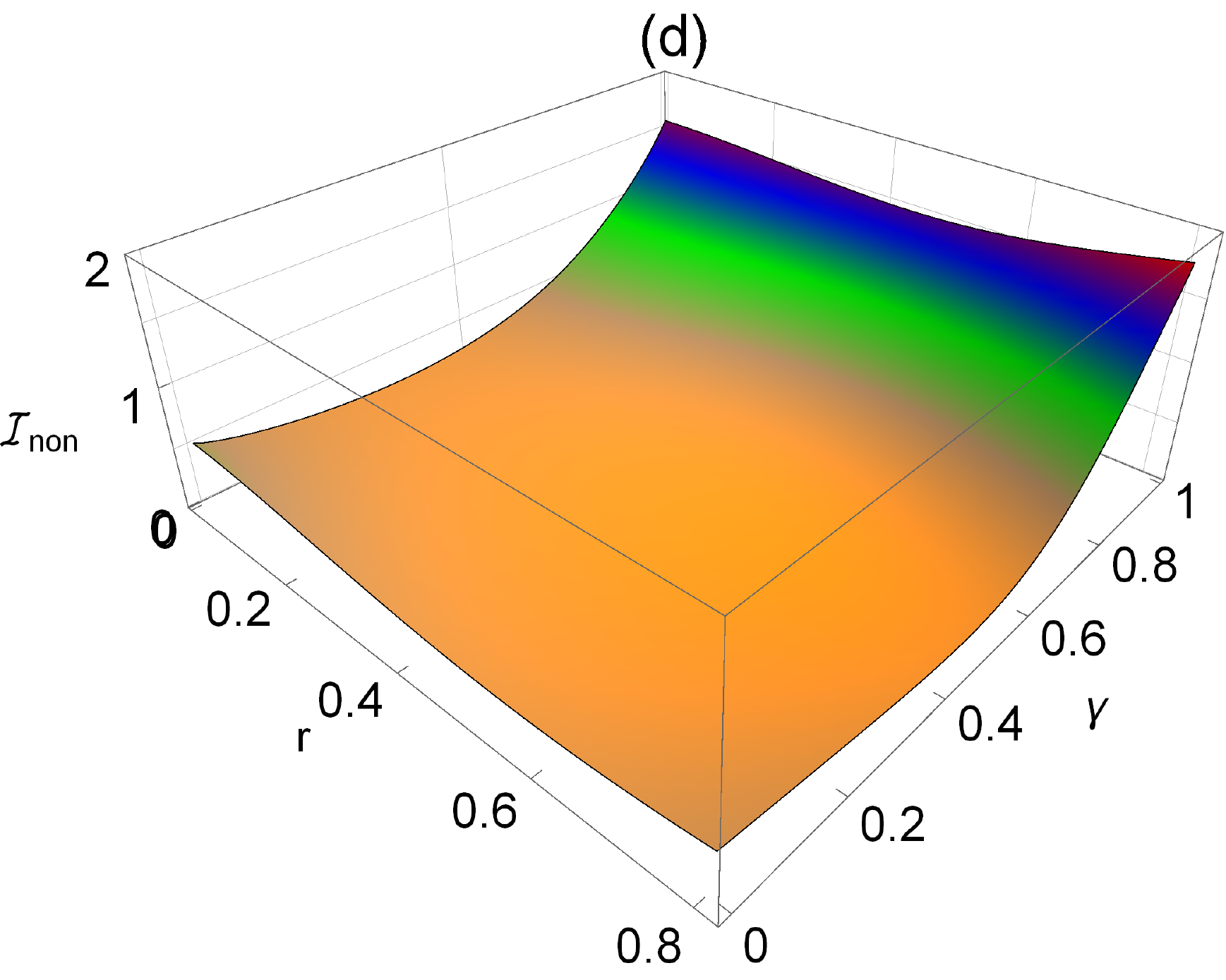}
	\caption{The same as Fig.(\ref{fig4}) but for the  global dephasing channel.}
	\label{fig6}
\end{figure}

Fig.(\ref{fig6}) shows the effect of the global dephasing channel on the degree of coherence and the non-local information that coded on the initial accelerated system. It is clear that, the behavior of the coherence that displayed in Fig.(\ref{fig6}a) is similar to that depicted in Fig.(\ref{fig4}a) for the multi-global dephasing channel. On the other hand, there is a dramatic effect of the global dephasing channel on the non-local information. As it is  displayed from   Fig.(\ref{fig6}b), the non-local information $\mathcal{N}_{non}$ increases gradually for non-accelerated system, namely at $r=0$. Moreover, the maximum values of $\mathcal{I}_{non}$ are predicted at maximum values of $r$ and $\gamma$, while the minimum values of the non-local information are displayed  when $r\to\infty$. This means that the possibility of keeping the non-local information on the accelerated system increases if  one switches on the global dephasing  channel.

For the system that is initially prepared in a bound entangled state, the behavior of the coherence $\mathcal{R}_e$, and the non-local information $\mathcal{I}_{non}$ is displayed in Figs.(\ref{fig6}c) and (\ref{fig6}d). The general behavior is similar to that displaced for free entangled  qutrits system, but the upper bounds are smaller than those for free entangled system.

\subsubsection{Amplitude Damping Channel:}
 In this subsection, we are forced the accelerated system (\ref{acc}) to interact locally with multi-local or global amplitude damping decoherence channel, where the Kraus super-operators of this decoherence channel are defined by:
 \begin{equation}\label{2}
 	\begin{split}
 		&\mathcal{E}_1=diag(1,\sqrt{1-\gamma},\sqrt{1-\gamma}) , \quad	\mathcal{E}_2= \sqrt{\gamma}|0\rangle \langle 1| , \quad	\mathcal{E}_3= \sqrt{\gamma}|0\rangle \langle 2|,
 	\end{split}
 \end{equation}

\begin{figure}[h!]
	\centering
	\includegraphics[width=0.42\linewidth, height=4cm]{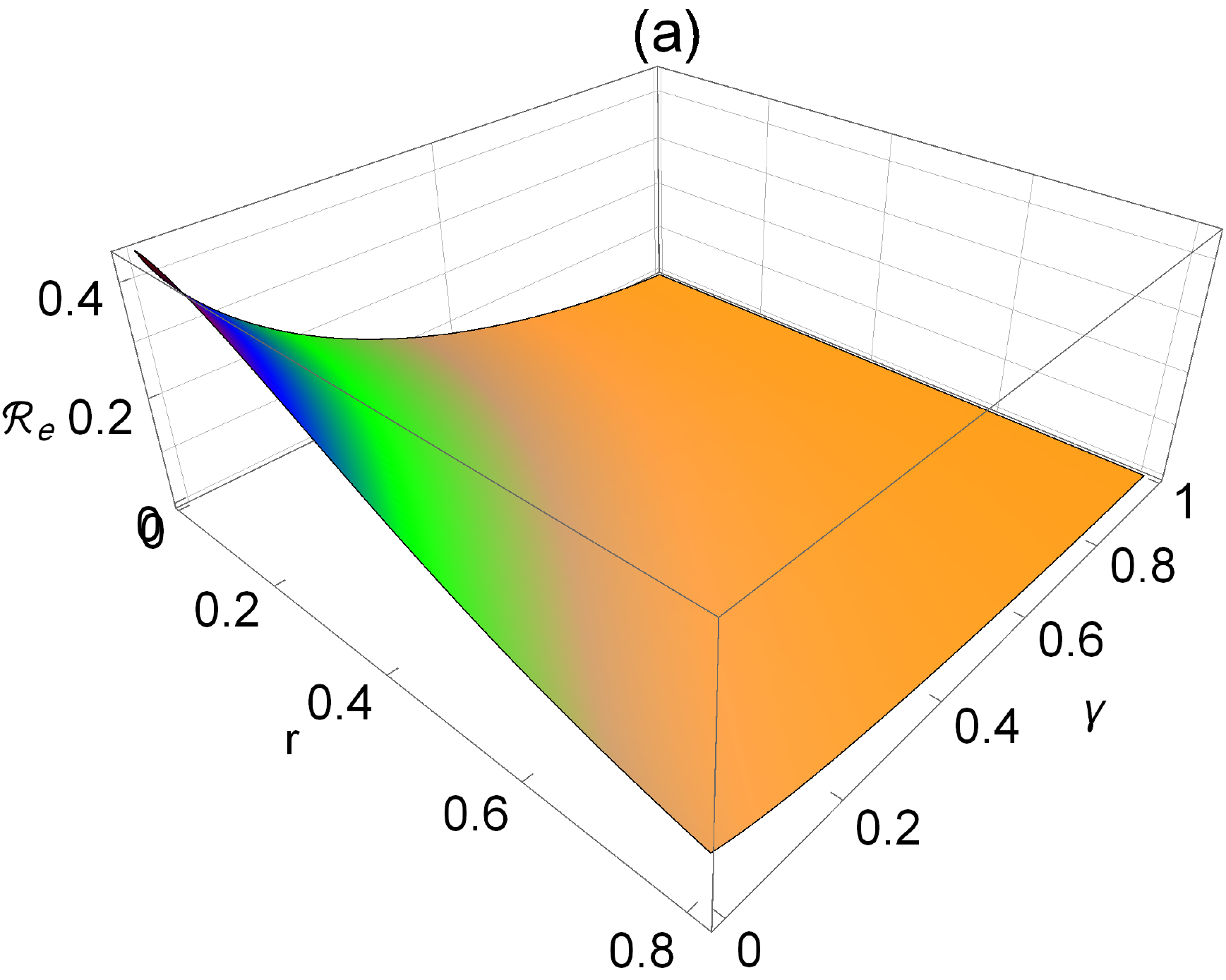} \hspace{0.2cm}
	\includegraphics[width=0.42\linewidth, height=4cm]{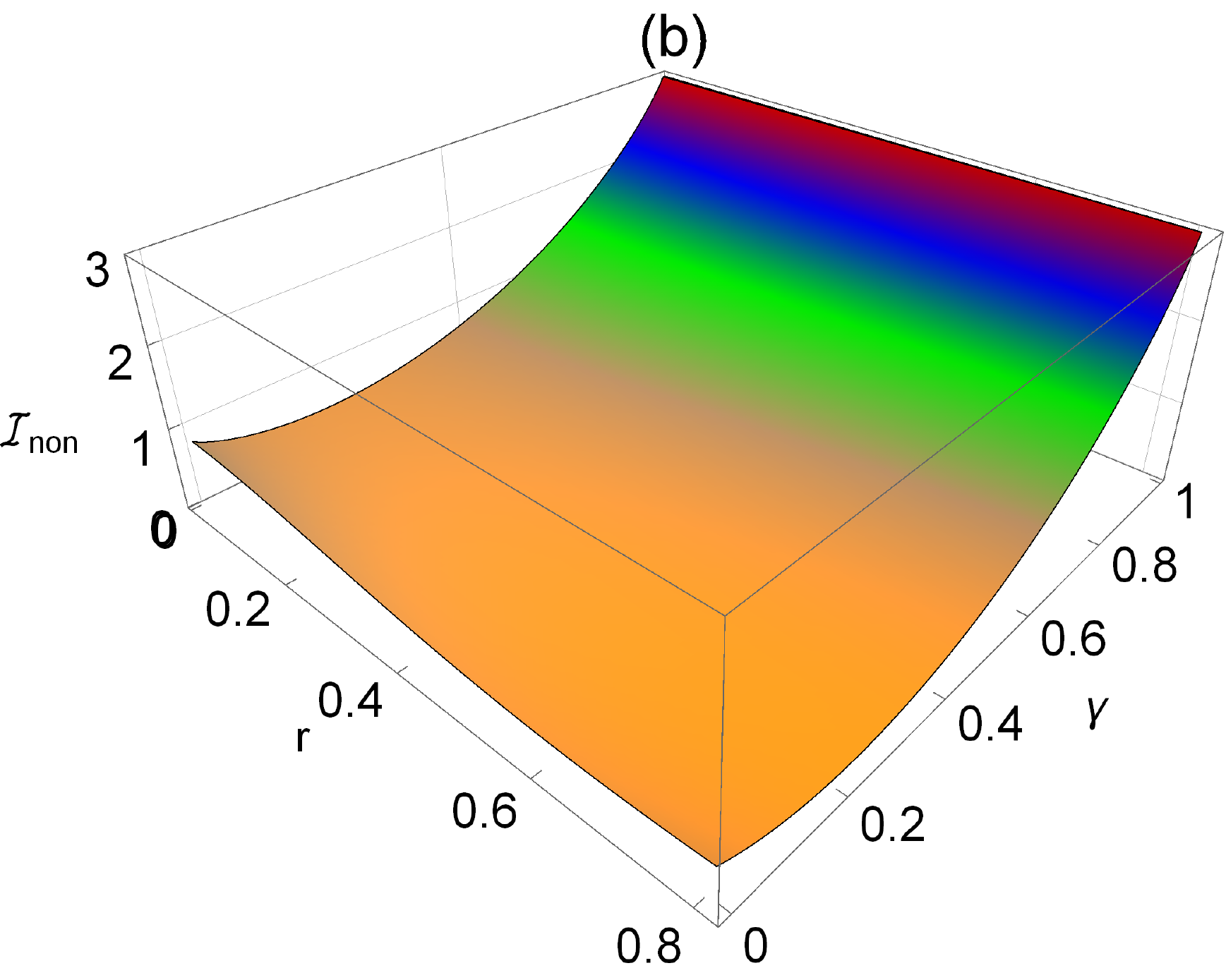}
	\includegraphics[width=0.42\linewidth, height=4cm]{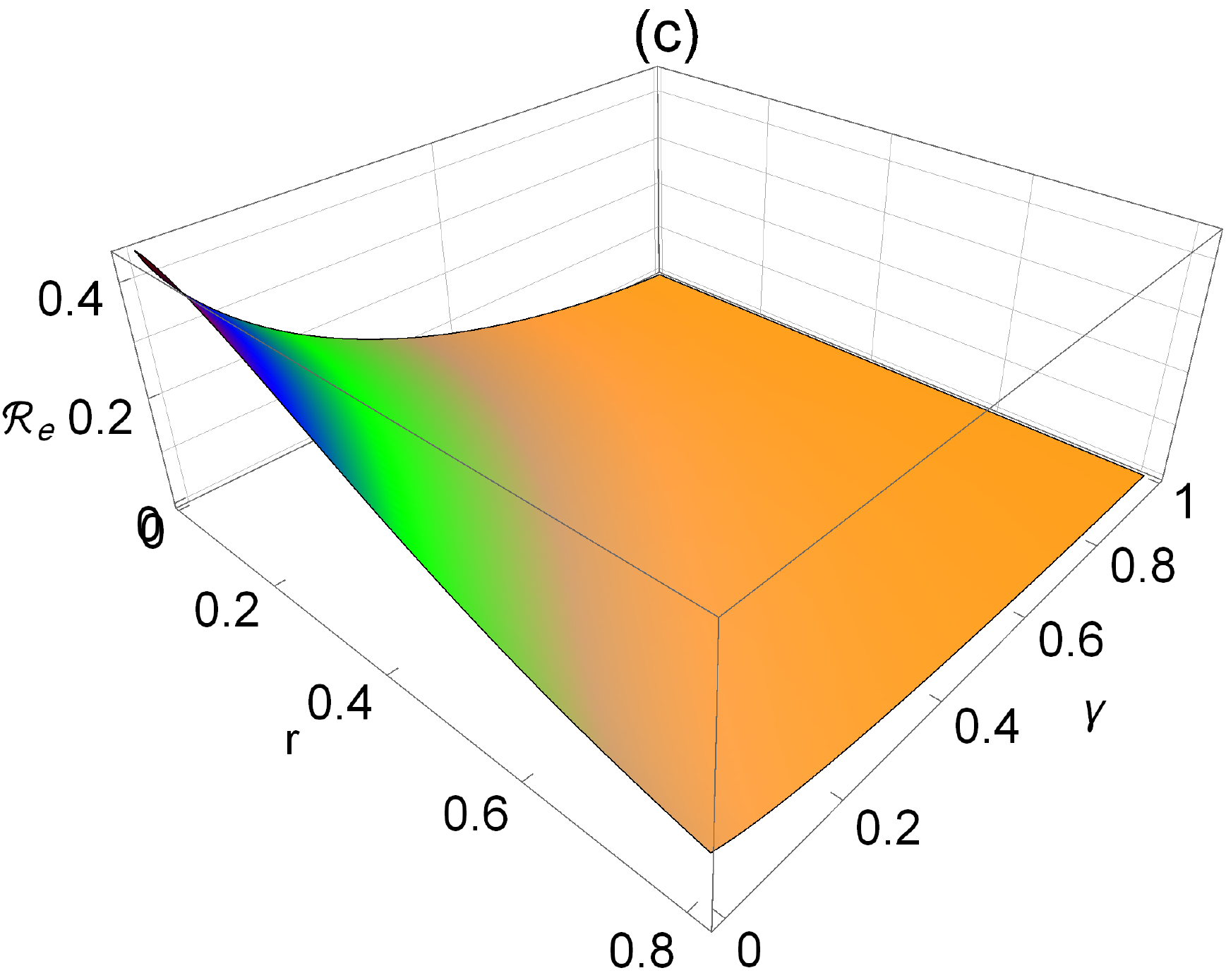} \hspace{0.2cm}
	\includegraphics[width=0.42\linewidth, height=4cm]{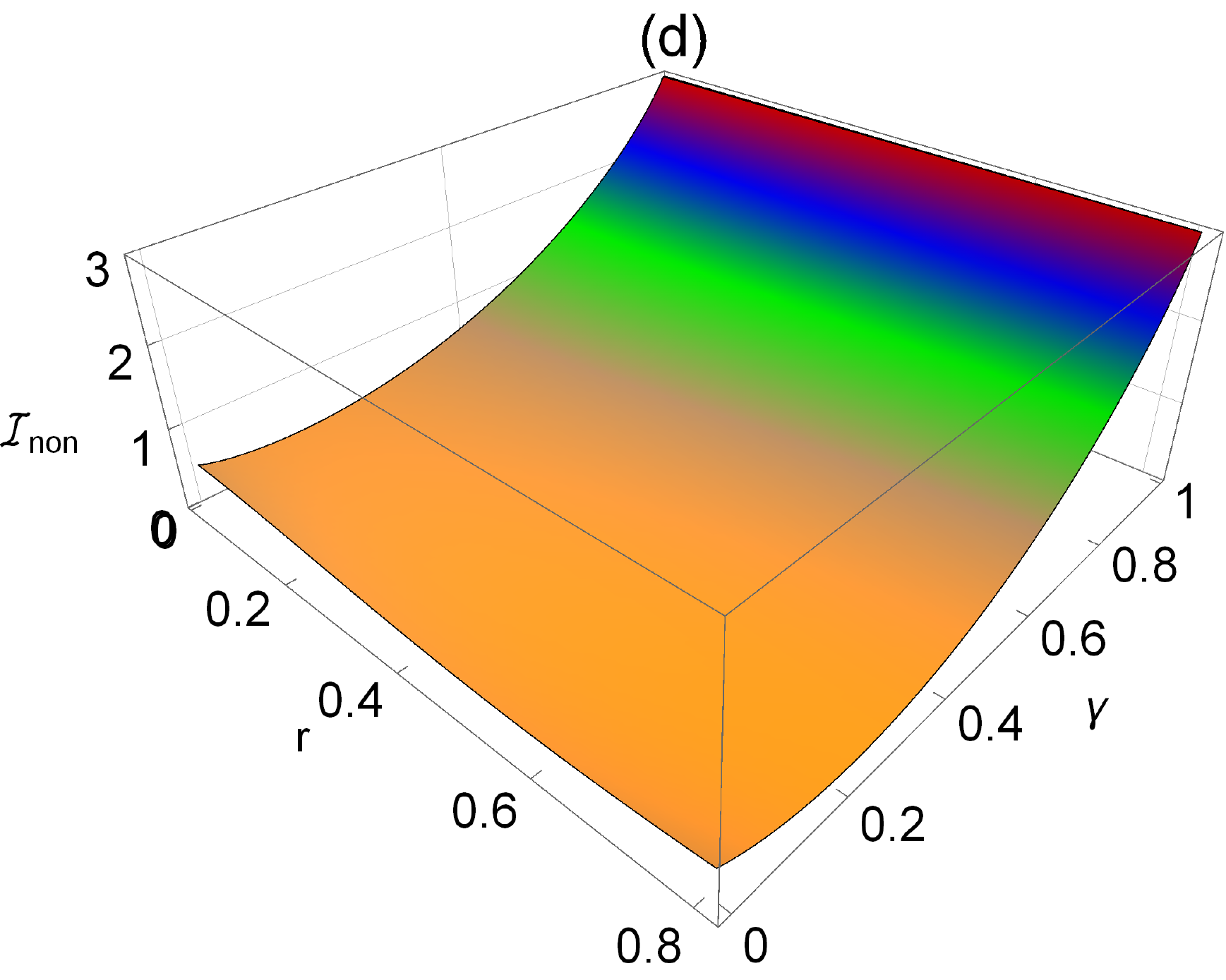}
	\caption{The same as Fig.(\ref{fig4}), but the accelerated systems passes through the multi-amplitude local   damping channel.}
	\label{fig7}
\end{figure}

In Fig.(\ref{fig7}), we  examine the effect of the multi-amplitude  damping channels on the behavior of the degree of coherence and the amount of the non-local information, where it is assumed that initial system is prepared in the free entangled interval, namely we set $\alpha=4.5$. As it is displayed from  Fig.(\ref{fig7}a), the behavior of $\mathcal{R}_e$  is similar to that displayed  for the dephasing channel.  On the other hand, the amount of non-local information $\mathcal{I}_{non}$ is larger than those displayed in Fig.(\ref{fig4}b). Moreover, the minimum values of the non-local information for the amplitude damping channel  are predicted at $r\to\infty$ and zero strength.
 Additionally, the maximum values of the non-local information are displayed as one increases the acceleration and the channel parameter $\gamma$.

\begin{figure}[h!]
	\centering
	\includegraphics[width=0.42\linewidth, height=4cm]{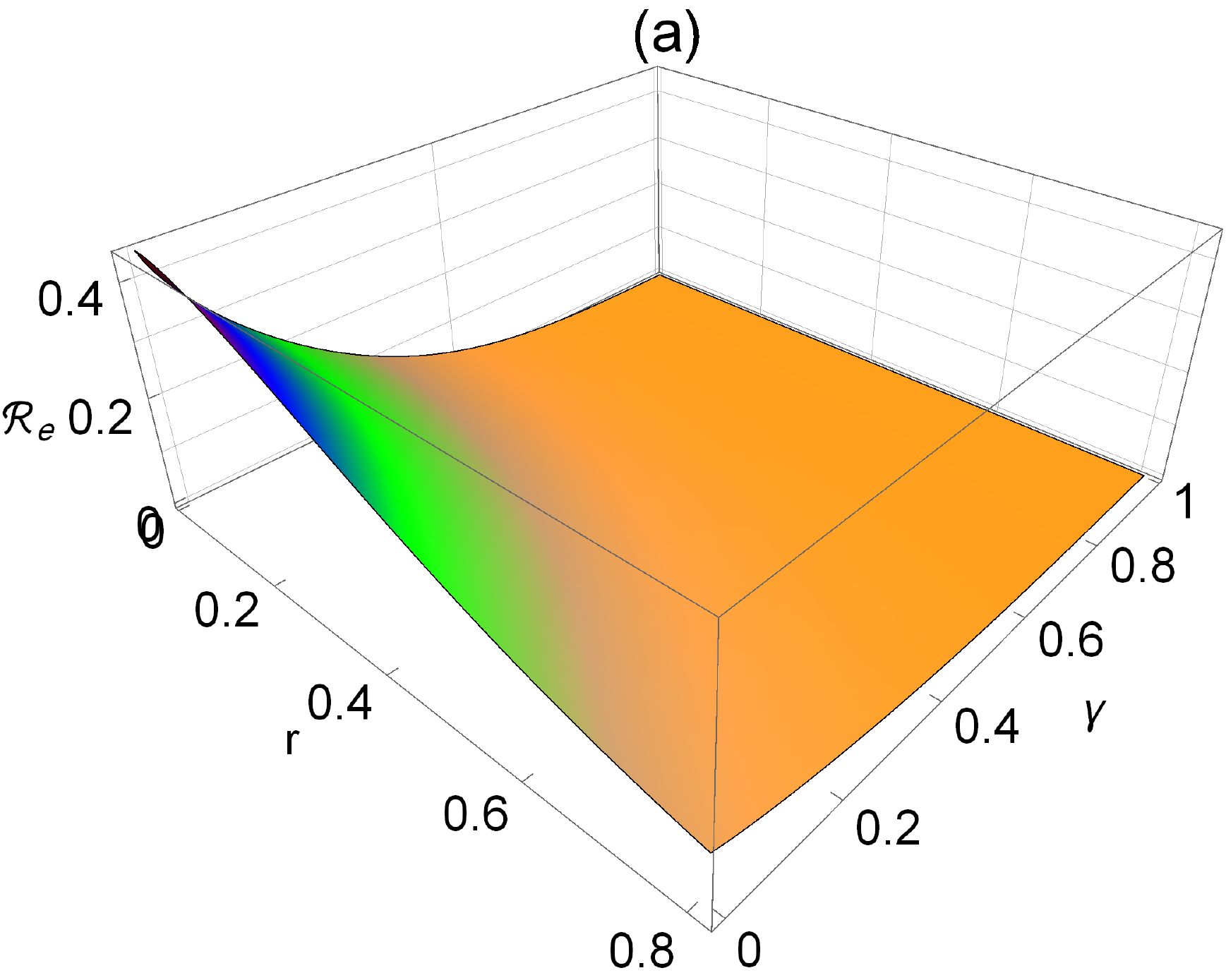} \hspace{0.2cm}
	\includegraphics[width=0.42\linewidth, height=4cm]{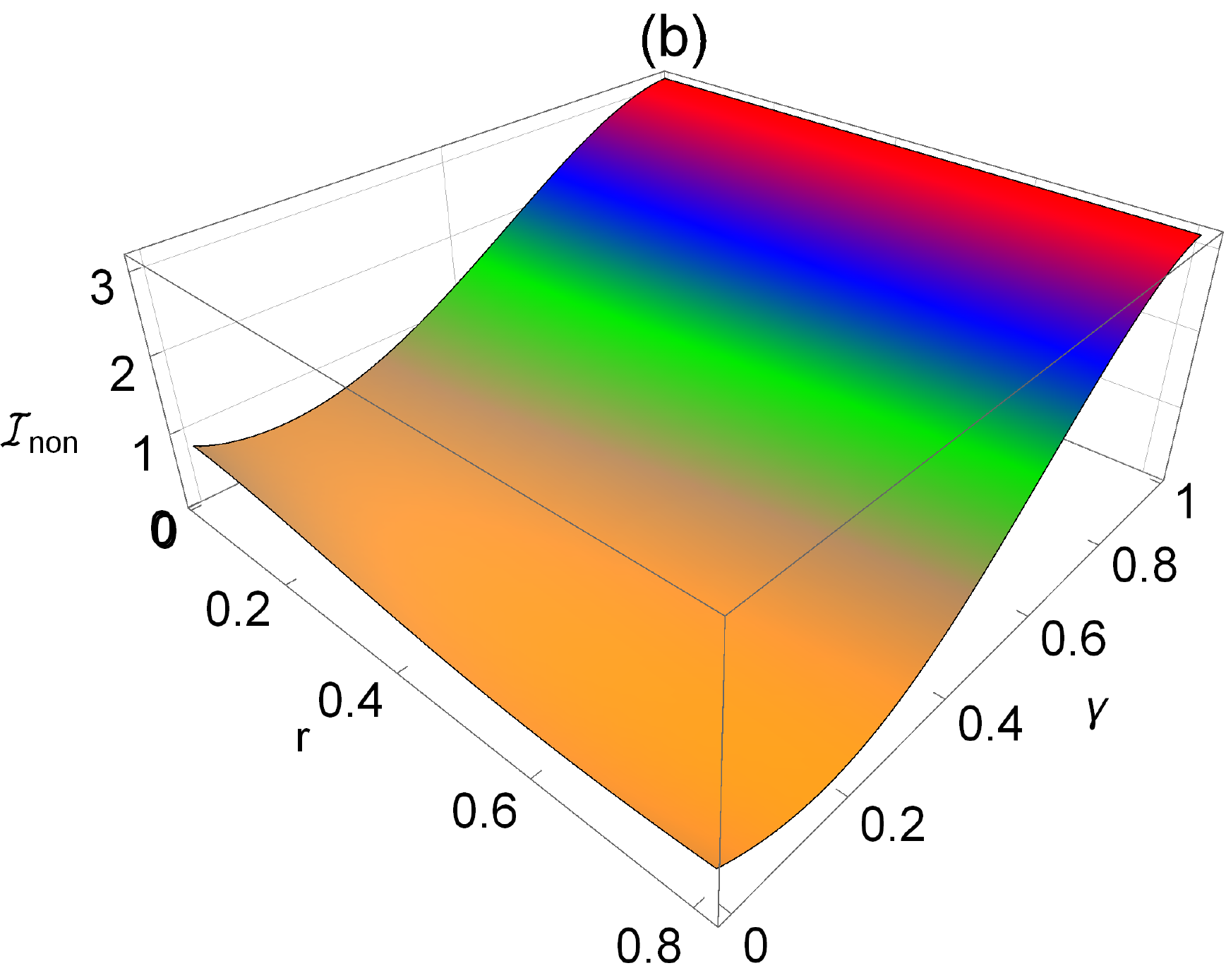}
	\includegraphics[width=0.42\linewidth, height=4cm]{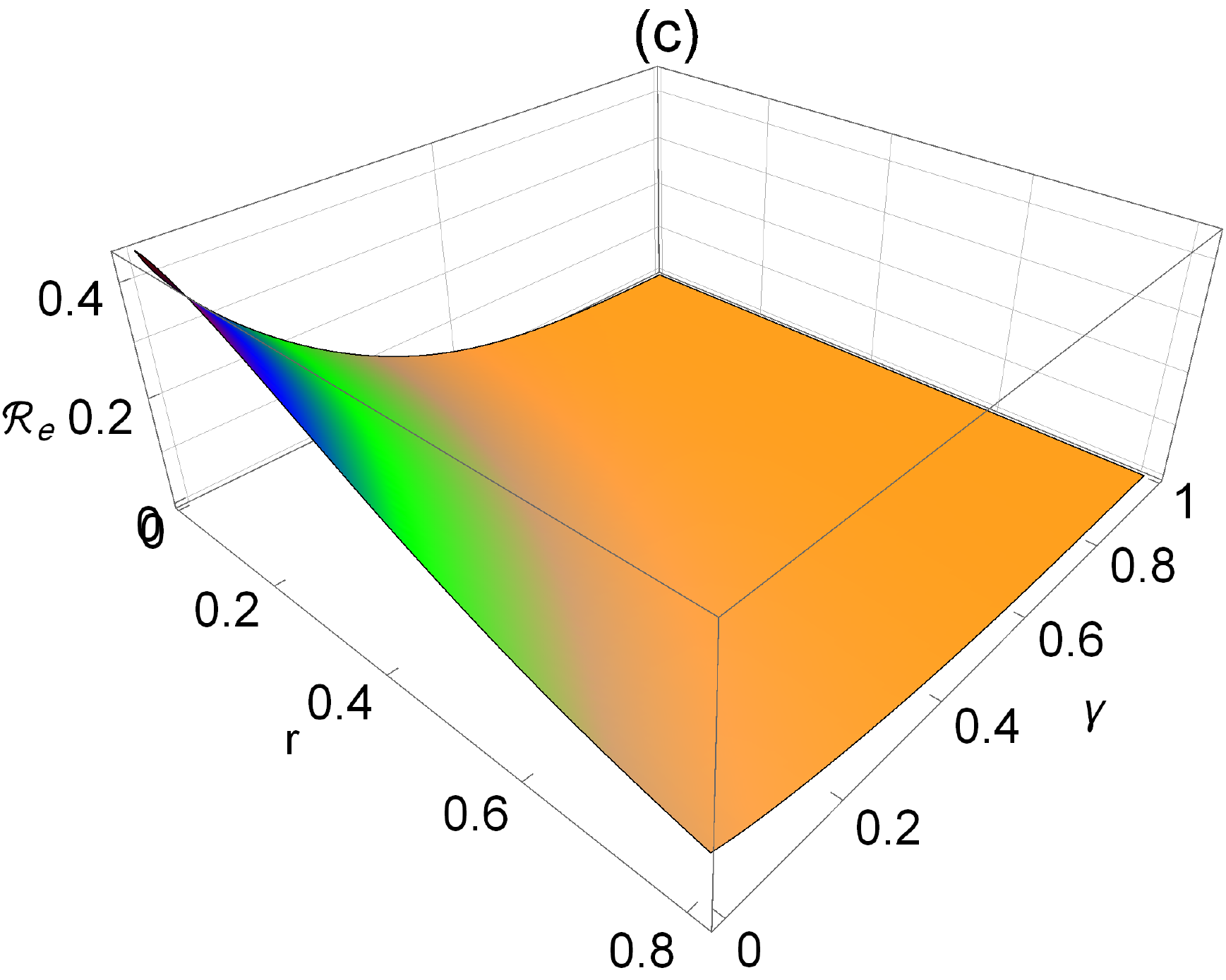} \hspace{0.2cm}
	\includegraphics[width=0.42\linewidth, height=4cm]{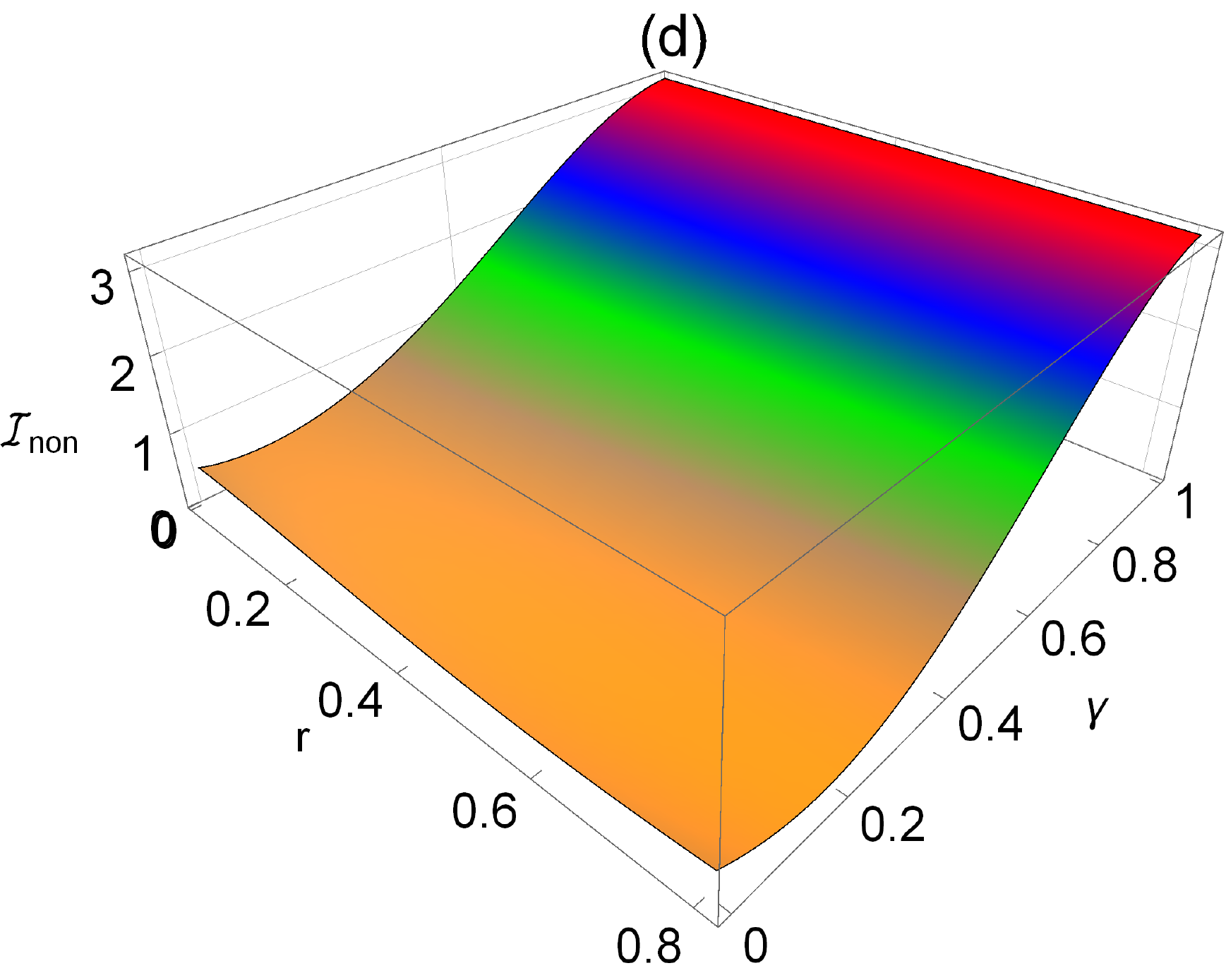}
	\caption{The same as Fig.(\ref{fig7}), but for the global amplitude damping channel.}
	\label{fig8}
\end{figure}

The effect of the global amplitude damping channel on $\mathcal{R}_e$ and $\mathcal{I}_{non}$ is displayed in Fig.(\ref{fig8}), where it is assumed that the accelerated system is initially prepared on bound entanglement's interval. The behavior of the  $\mathcal{R}_e$ is similar to that displayed in the previous figures. However, for the non-local information the global amplitude damping channel has a constructive behavior, where $\mathcal{I}_{non}$ increases as one increases the channel strength $\gamma$. Moreover, it decays slightly as one increases the acceleration.
From Figs.(\ref{fig7}b,d) and (\ref{fig8}b,d), one may conclude that the ability of the   global amplitude damping channel to protect the amount of the non-local information is much better than that displayed for the multi-local dephasing channel. Moreover, the information which passes through the amplitude noisy channel is more robust than that passes through the dephasing channel.

\section{Conclusion}\label{s5}
\qquad In this manuscript, we investigate the effect of local and global noisy dephasing and amplitude channels on the behaviour of entanglement, coherence and non-local information of the accelerated two-qutrit system. It is assumed that the initial qutrit system behaves as entangle or bound entangled system. The effect of the acceleration parameter and the strengths of the noisy channels are discussed.

The obtained results showed that, in the absence of the noisy channel, the three phenomena; entanglement, coherence and non-local information are decreased gradually as the acceleration increases. On the other hand, the initial state settings have an impact effect on both quantities, where they increase as the weight parameter increases. More precisely, the amount of coherence and the non-local information that is depicted for the free initial entangled systems are more robust than that displayed for the initial bound entangled system.

The effect of the noisy channels on the behaviour of the coherence’s degree and the non-local information of the accelerated initial system is discussed. Due to the external noise, both phenomena are affected, where the upper bounds of the coherence’s degree decrease as the strengths of the channels increase. The decreasing rate depends on the type of noisy channel, whether it is dephasing or an amplitude, where the non-local information increase as the strengths of both channels are increased. However, the increasing rate of the non-local information because of the amplitude damping channel is larger than that displayed for the dephasing channel. The behaviour of the coherence shows that it decreases gradually as the acceleration or the channel strengths are increased.

In this context, it’s worth noting an observation that, for the multi-local noisy channels, the acceleration has an observable effect on the behaviour of the non-local information. Meanwhile, this effect may slightly appear in the presence of the global noisy channels. The physical reason behind this behaviour is that both particles are accelerated local. Therefore, there is an additional effect due to the acceleration process.

The initial state settings have a noticeable effect on the behaviours of the coherence and the non-local information of the accelerated qutrits system. Coherence decreases gradually if the initial qutrits system is prepared in the free or bound entangled intervals. However, the decreasing rate displayed for the system initially prepared in a free entangled interval is smaller than that prepared in the bound entangled interval. The coherence decreases gradually if the system passes through the multi-local noisy channels, while sharply in the presence of the global noisy channels. The got numerical results show that the predicting increasing rate of the non-local information for systems prepared initially in a free entangled interval is larger than that displayed for systems prepared in the bound entangled interval.

{\it In conclusion}, the amount of coherence and the non-local information of the accelerated two-qutrit system could be protected by passing it in dephasing and amplitude damping channels. This process decreases the decoherence that arises from the acceleration process. The initial state settings have an impact on the possibility of protecting the physical quantises that may lose their efficiency during the acceleration process.

\end{document}